\documentclass[lettersize,journal]{IEEEtran}
\usepackage{amsmath,amsfonts}
\usepackage{algorithmic}
\usepackage{algorithm}
\usepackage{array}
\usepackage[caption=false,font=normalsize,labelfont=sf,textfont=sf]{subfig}
\usepackage{textcomp}
\usepackage{stfloats}
\usepackage{url}
\usepackage{verbatim}
\usepackage{graphicx}
\usepackage{tabularx}
\usepackage{cite}
\usepackage{booktabs} 
\usepackage{pifont} 
\usepackage[table]{xcolor} 
\usepackage{multirow}
\usepackage{lipsum}
\usepackage{makecell}
\usepackage{enumitem}
\definecolor{revisionblue}{RGB}{0,102,204}

\begin{document}

\title{Which Deep Learner? A Systematic Evaluation of Advanced Deep Forecasting Models Accuracy and Efficiency for Network Traffic Prediction.}

\author{Eilaf MA Babai,~\IEEEmembership{Student Member,~IEEE,} Aalaa MA Babai,~\IEEEmembership{Student Member,~IEEE,} Okamura Koji,~\IEEEmembership{Member,~IEEE,}
\thanks{Kyushu University, Fukuoka, Japan (babai.eilaf.756@s.kyushu-u.ac.jp; babai.aalaa.103@s.kyushu-u.ac.jp; oka@ec.kyushu-u.ac.jp).}}

\maketitle

\begin{abstract}
Network traffic prediction is an essential tool for automating modern network management. It is a difficult time series forecasting (TSF) problem that has been addressed by Deep Learning (DL) models due to their ability to capture complex patterns. Recent years have witnessed advances in deep forecasting models, from sophisticated transformer architectures to simple linear models, improving performance across diverse prediction tasks. However, given the diversity of these models, the variability of network traffic across environments and timescales, and the inherent challenges of network traffic prediction, it is essential to identify effective deployment choices and modeling directions. 
Henceforth, this study systematically identifies 12 deep forecasting models, comprising 8 advanced TSF models and 4 traditional deep learning architectures -each with unique advantages for network traffic prediction- against 3 statistical baselines, for a total of 15 forecasting methods and evaluate them on four real traffic datasets, across multiple time scales and horizons, assessing performance, robustness to anomalies, data gaps, external factors,  data efficiency, and resource efficiency in terms of time, memory, and energy. Our results provide insights into model behavior and guidelines for network management, highlighting performance regimes, efficiency thresholds, and promising architectures that balance accuracy, efficiency, and robustness to traffic challenges.
\end{abstract}


\section{Introduction}

\IEEEPARstart{M}odern network management relies on network traffic prediction to optimize and automate various tasks. For example, based on estimated future traffic, network resources can be allocated in advance to avoid congestion\cite{9141210}. Malicious attacks such as Denial-of-Service or irregular amounts of spam can be detected by comparing real traffic with the values predicted by forecasting algorithms\cite{krishnamurthy2003sketch}. Additionally, the energy consumption of underutilized devices can be reduced according to expected traffic values\cite{zhang2017traffic}. Thus, effective network traffic prediction enables proactive and cost-efficient network management. 

\textbf{The Network Traffic Prediction (NTP) Problem.} NTP problem is a time series forecasting problem in which various statistics of the network data flow are systematically captured and recorded by network devices. These data are then processed and aggregated at discrete time intervals to generate a network traffic time series suitable for analysis and prediction. The primary goal is to develop a reliable forecasting model that accurately estimates future network traffic and can be seamlessly integrated into existing network infrastructure for operational deployment. 

Prediction accuracy affects network management; large errors can cause underprovisioning, or resource wastage, and in anomaly detection, frequent errors lead to frequent false alarms\cite{8076830}. These situations result in service disruptions and financial losses.  
Meanwhile, the operational deployment of the model stresses resource efficiency. 
Network devices typically have limited processing and memory resources, and upgrades can be complex and costly \cite{9762358}. Furthermore, due to the dynamic nature of networks, maintaining model performance often requires frequent retraining, incurring substantial energy costs. Hence, the NTP field focuses on developing accurate and efficient prediction models.

\textbf{NTP Methods.} 
Current NTP methods involve:   (1) Traditional statistical techniques, (2) Deep learning (DL), and (3)  Advanced time series forecasting models (TSF).
Statistical methods are simple but limited to linear modeling of temporal relationships. Thus, they fail to capture complex traffic patterns. In contrast, DL models' ability to learn intricate nonlinear relationships from large datasets makes them viable for NTP\cite{ferreira2023forecasting, aouedi2025deep}. For instance, the Multilayer Perceptron (MLP) outperforms statistical methods in 5-minute and 1-hour traffic forecasting with shorter training and testing times\cite{cortez2006internet}. Recurrent Neural Network (RNN) architectures achieve good performance in predicting traffic volume, packet protocols, and packet distributions \cite{ramakrishnan2018network}. Temporal Convolutional Networks (TCN) successfully capture short-term features from network web traffic \cite{bi2021hybrid}, while Transformer models effectively capture temporal traffic features \cite{cityleveltraffic} and overall network dynamics \cite{dietmuller2022new}. On the other hand, TSF models emerged from recent advances in time series forecasting. For instance, Transformer-based TSF models, such as PatchTST\cite{nie2022time} and Autoformer\cite{Autoformer2021}, employ innovative representation techniques, such as patching and time-domain decomposition, and have achieved state-of-the-art performance in forecasting. There is also a growing focus on models that incorporate exogenous variables, such as the Temporal Fusion Transformer (TFT) \cite{TFT2021}, TiDE \cite{das2023long}, and TimeXer \cite{wang2024timexer}, which demonstrate improved forecasting performance with external covariates. From hereafter, we refer to TSF and DL collectively as \textit{deep forecasting models.}

\textbf{Real-world NTP Accuracy-Efficiency Requirements and Inherent Traffic Challenges Raise Important Questions}.
Nevertheless, despite the plethora of deep forecasting models, no single model consistently outperforms others across all traffic prediction scales \cite{qiao2022deep} or network environments \cite{koumar2025comparative}, indicating significant potential for further advances toward generalizable models. Aiding the informed development of such models makes it pertinent to first address \textit{\textbf{RQ1:} Which forecasting models exhibit the strongest generalization capabilities across NTP tasks while demonstrating robustness to challenging traffic characteristics?}

Crucially, the rapid growth of deep forecasting models
 has also been met with scrutiny regarding their actual performance. Makridakis et al.\cite{makridakis2022m5} reported that many DL approaches can be outperformed by simple statistical baselines like Exponential Smoothing. Furthermore, Zeng et al. \cite{transformer_effective_2023} show that advanced transformer-based models may be surpassed by a simple linear layer combined with decomposition techniques. These findings underscore significant questions for the deployment and future exploration of deep forecasting models for NTP,  as emphasis grows on model performance and efficiency in data and resource use. Thus, we must address both
\textit{\textbf{RQ2:} How do various network traffic forecasting models compare in terms of data efficiency, particularly in maintaining prediction accuracy when trained with limited datasets?} and \textit{\textbf{RQ3:} Which forecasting models offer the highest resource efficiency when deployed in low-resource network environments?}

Addressing these questions calls for a systematic evaluation of complex models for multi-scale consistency and robustness (RQ1),  data efficiency (RQ2), and resource efficiency (RQ3).

\textbf{Related Work in Evaluation NTP Robustness and Efficiency is Limited to Prevalent Deep Learning Architectures.}
Several studies have empirically compared the performance of various deep learning models in network traffic prediction 
\cite{saha2023outliers, ferreira2023forecasting, qiao2022deep, aouedi2025deep}. However, their focus is on limited deep learning architectures and does not cover advanced time series forecasting models. Moreover, while some studies mention challenging network traffic characteristics \cite{aouedi2025deep} and deployment challenges \cite{ferreira2023forecasting}, they lack analysis of how these challenges affect model performance. 
Finally, there is a lack of comprehensive evaluation of model size, training time, and energy consumption, which are critical deployment considerations given the frequent need for retraining. We identify gaps in the literature (Table \ref{tab:related_work}) and emphasize the need for a thorough evaluation of deep forecasting models under real-world traffic challenges and for assessing their efficiency.

\textbf{Our Primary Objective and Approach.} 
Our primary objective is to address these gaps, distilled in our aforementioned questions RQ1, RQ2, and RQ3.
To that end, we systematically evaluate recent paradigms in time series forecasting and prevalent deep learning architectures within the context of NTP challenges, investigating how their advantages, shortcomings, and suitability change across NTP tasks and deployment conditions. Thereby, we provide a benchmark and a guide for future model development and deployment. 
\textbf{First,} we characterize the unique challenges of the network traffic time series across \(4\) real traffic datasets at multiple time scales. These are outliers, missing values, and the influence of external factors. 
\textbf{Second,} we implement \(12\) deep forecasting models, comprising \(8\) advanced TSF models and \(4\) DL architectures, together with \(3\) baselines. 
Advanced TSF models are relatively new and are not readily available in standard time series forecasting libraries. These models also have numerous distinct hyperparameters, complicating direct comparison without a comprehensive understanding of each model.
\textbf{Third,} we evaluate the generalization of these models across multiple traffic time scales, their robustness to challenges in traffic data quality, their data efficiency with limited data, and their resource efficiency for deployment.

In summary, our contributions are:
\begin{itemize}
    \item \textbf{Identifying and Evaluating TSF Trends} (Section \ref{sec:methodology}): We select \(8\) advanced TSF models with advantageous characteristics for NTP, discuss their innovations, and evaluate them together with 4 traditional DL architectures, forming a set of \(12\) diverse deep forecasting models evaluated on multi-scale and multi-horizon NTP tasks.
    \item \textbf{Developing a Systematic Methodology and Comprehensive Benchmarking} (Section \ref{sec:methodology}):
    We evaluate models on isolated traffic data challenges—anomalies, missing data, and external variations—across various network environments and timescales, yielding multiple versions of \(11\) real-world network datasets. Additionally, we assess deployment efficiency by measuring data and resource utilization, including energy consumption, model size, and training duration, while varying data size. We intend to release our model implementations and experimental framework upon publication. 
    \item \textbf{Insights and Guidelines:}(Sections \ref{sec:exp_generalization&Robustness}--\ref{sec:which_deep_learner}) By addressing \textit{RQ1}-\textit{RQ3}, we derive multiple insights that characterize deep forecasting model performance and efficiency for network traffic prediction, culminating in practical model selection guidelines; the key findings are summarized as follows:
\end{itemize}

\noindent 
\textbf{Insight 1.} Model selection strongly depends on the traffic timescale. Pareto-optimal NTP models shift from sequential, linear, and compact MLP models at fine scales towards patch-based transformers and MLP encoder--decoders at medium and hourly scales, where recurring traffic structure becomes more pronounced. These differences lead to distinct accuracy--resource trade-offs across NTP tasks.\\
\textbf{Insight 2.} The distribution of missing telemetry matter more than its percentage. Under the tested zero-filled conditions, short distributed gaps cause substantially greater degradation than longer concentrated gaps, highlighting the importance of preserving local traffic continuity.\\
\textbf{Insight 3.} Data requirements differ substantially across forecasting architectures, with three weeks as the empirical \textit{information quantity} threshold for data-efficient NTP models. DLinear, TiDE, and patch-based models retain more of their accuracy with reduced history than sequential models and NBeats. The data-efficient group incurs about 6\% higher error with three weeks of history than with twelve, supporting shorter histories for frequent retraining.\\
\textbf{Insight 4.} DLinear remains Pareto-optimal in 83\% of the evaluated dataset–resource combinations. Its combination of forecasting accuracy, low resource costs, data efficiency, and robustness makes it a strong initial candidate for resource-constrained deployment.\\
\textbf{Insight 5.} External covariates provide selective benefits and do not consistently resolve recovery from abnormal traffic periods. Holiday- and weather-related capacity and energy schedules require evaluation of event-window accuracy and recovery-period errors alongside overall forecasting performance.\\
\textbf{Insight 6.} Tuning benefits and costs differ substantially across models. PatchTST and TimeXer achieve average tuning gains of approximately 17\% and 18\%, while DLinear shows limited reliance on task-specific tuning. Operators should consider both achievable improvements and the effort required to obtain them when adapting predictors across traffic settings.\\
\textbf{Insight 7.} Strong accuracy on clean traffic does not ensure robustness to outliers. DLinear and TiDE exhibit lower degradation under the tested outlier conditions, showing why telemetry quality and outlier handling should influence model selection alongside ideal-data accuracy.\\

\begin{table*}[t!]
\centering
\caption{Comparison between our study and related comparative studies on deep learning models for network traffic prediction. We provide a comprehensive review of advanced forecasting models over multiple network traffic time scales. A present \ding{51} indicates an addressed category, while a missing \ding{51} indicates it is not being addressed by the study. }

\label{tab:related_work}
\renewcommand{\arraystretch}{1.2}  
\setlength{\tabcolsep}{3pt}  
\begin{tabularx}{\textwidth}{ l  p{0.055\linewidth}  c  c  >{\centering\arraybackslash}X >{\centering\arraybackslash}X >{\centering\arraybackslash}X c  c  c  c  >{\centering\arraybackslash}X  c  c  c }
\toprule
    \multirow{3}{*}{\textbf{Work}}  & \multicolumn{2}{c}{\textbf{NTP Task}}
    & \multicolumn{3}{c}{\textbf{Data Quality}}& \textbf{Data Quantity} & \multicolumn{4}{c}{\textbf{Deep Learning Models}}
    & \textbf{TSF Models} & \multicolumn{3}{c}{\textbf{Model Efficiency}}\\
    \cmidrule(l){2-3}
    \cmidrule(l){4-6}
    \cmidrule(l){8-11}
    \cmidrule(l){13-15}

 & Time Scale & Horizon & Outliers &Missing Values& External Factors &  & MLP & RNN & CNN & Transf. &  &Size& Time & Energy \\
\midrule

Oliveira et al. (2016)\cite{oliveira2016computer} & M, H, D & Short &  && & & \ding{51} & \ding{51} & & & & & \ding{51} & \\

Jiang (2022) \cite{jiang2022internet} & H & Short &  &&  & & \ding{51} & \ding{51} & \ding{51}& \ding{51}& &\ding{51}& \ding{51} & \\


Ferreira et al. (2023) \cite{ferreira2023forecasting} &H &Short &  &&  & & & \ding{51} & \ding{51} & & &\ding{51}&  & \\

Saha et al. (2023) \cite{saha2023outliers} & M &Short & \ding{51}  &&  & & & \ding{51} & & & &&  & \\

Aouedi et al. (2025)\cite{aouedi2025deep} & M &Short &  &&  & & & \ding{51} & \ding{51} & & && \ding{51} & \\

Koumar et al. (2025)\cite{koumar2025comparative} &H &Short &  &\ding{51}&  & & & \ding{51} & \ding{51}  & &\ding{51} && \ding{51} & \\

wang et al. (2024)\cite{wang2024deep}* & M, H, D & Short Long &  & &  & &  \ding{51}& \ding{51} &  \ding{51} &  \ding{51}&  \ding{51}&&  & \\

\midrule

\multirow{1}{*}{Our work} & M, H, D & \multirow{1}{*}{\makecell[c]{Short Long}} & \multirow{1}{*}{\ding{51}} & \multirow{1}{*}{\ding{51}}& \multirow{1}{*}{\ding{51}} & \multirow{1}{*}{\ding{51}}& \multirow{1}{*}{\ding{51}} & \multirow{1}{*}{\ding{51}}&  \multirow{1}{*}{\ding{51}}& \multirow{1}{*}{\ding{51}}& \multirow{1}{*}{\ding{51}}&\multirow{1}{*}{\ding{51}}& \multirow{1}{*}{\ding{51}}& \multirow{1}{*}{\ding{51}} \\

\bottomrule
\multicolumn{15}{l}{\makecell[l]{
M: Minutes, H: Hours, D: Days.
*Annotated work is in the general scope of time series forecasting and does not evaluate for network traffic data. }}

\end{tabularx}
\end{table*}

\IEEEpubid{0000--0000/00\$00.00~\copyright~2021 IEEE}

\section{Background}
\label{sec:background}

In this section, we discuss the challenging characteristics of the network traffic time series, formulate the network traffic prediction problem, and then review traditional forecasting methods serving as baselines.

\subsection{Characteristics of Network Traffic Time Series}
\label{sub_background:traffic_data_characteristic}
Network traffic data is systematically recorded using network flow collection protocols like NetFlow\cite{claise2004cisco}, which measure traffic passing through each network interface. This data is then aggregated into a time series reflecting traffic volume over relevant temporal scales, tailored to specific network management needs. Such time series exhibit unique statistical and temporal properties that present challenges for accurate forecasting. These properties are as follows.

\subsubsection{Periodicity and Non-stationarity}
Network traffic time series exhibit multiple periodic patterns, notably diurnal and weekly cycles that reflect user behavior\cite{lakhina2004structural}. The diurnal cycle causes increased daytime traffic and decreased nighttime traffic, while weekly patterns show lower activity on weekends. These seasonalities lead to non-stationarity where statistical properties like mean and variance vary based on the time of day and day of week, as illustrated in Fig. \ref{fig:background_trafficCharacteristics}(a). Therefore, network traffic predictors must effectively model these complex patterns.

\subsubsection{Irregular Fluctuations and Outliers}
The growing diversity of network services and traffic sources leads to irregular patterns and fluctuations, as shown in Fig. 
\ref{fig:background_trafficCharacteristics}(b). Furthermore, network traffic often exhibits outliers or anomalies caused by malicious activity, sudden shifts, or demand spikes for specific services
\cite{lakhina2004characterization}. Such anomalies impair performance and must be detected and addressed before prediction\cite{saha2023outliers}, which can be costly. A robust predictor needs to discern meaningful patterns from fluctuations and anomalies. Fig. 
\ref{fig:background_trafficCharacteristics}(c) illustrates an example of an anomaly.

\subsubsection{Prolonged Data Gaps Due to Collection Challenges} 
Reliable collection of network traffic is difficult due to the high volume and increasing complexity of modern networks. Collection devices are susceptible to overloads and malfunctions, causing extended data gaps that hinder interpolation and impair the learning of network dynamics. Therefore, robust traffic prediction models must remain accurate despite substantial missing data. Fig. \ref{fig:background_trafficCharacteristics}(d) illustrates a three-day period with missing traffic data.

\subsubsection{Variations and Influence of External Factors}
\begin{figure*}
    \centering
    \includegraphics[width=\linewidth]{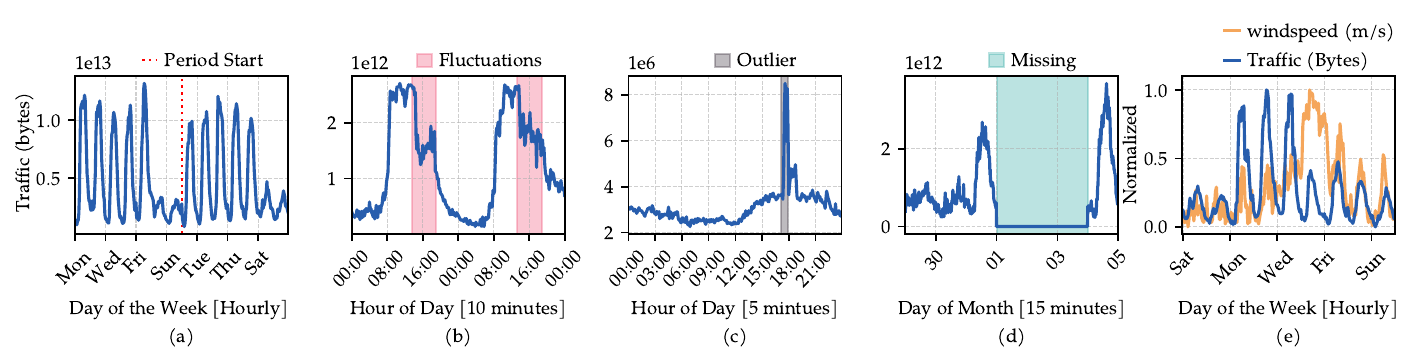}
    \caption{Characteristics of network traffic time series across multiple time scales. (a) Multi-Periodicity at the hourly scale. (b) Fluctuations at the 10-minute scale. (c) Anomalies at the 5-minute scale. (d) Missing data on the 15-minute scale. (e) Variations due to External factors (windspeed\cite{windspeed}) at the hourly scale.}
    \label{fig:background_trafficCharacteristics}
\end{figure*}

Traffic time series display temporal variations caused by external factors such as holiday \cite{xu2018hybrid}, weather \cite{padmanabhan2019residential}, and social events \cite{pimpinella2022using, bejarano2021deep, babai2024contextual}. These factors may induce increases or decreases in traffic, depending on the network context. These variations can be predicted using covariates or exogenous variables—auxiliary time series that estimate the timing and duration of external influences—as shown in Fig. \ref{fig:background_trafficCharacteristics}(d). Reliable traffic predictors must effectively account for external factors.

\subsubsection{Varying dynamics by time scale}
While the discussed characteristics are present across all traffic timescales, their prominence varies. Periodic patterns dominate at larger scales, such as hourly and daily intervals, which have smoother patterns. Conversely, fluctuations and anomalies are more common at finer, minute-level scales. Missing data disproportionately affects smaller aggregations, creating larger gaps. Different timescales correspond to specific networking tasks: capacity planning and energy management require longer intervals (hours), whereas resource allocation and anomaly detection rely on shorter ones.
These different dynamics indicate that prediction methods, whether statistical or deep learning\cite{qiao2022deep}, exhibit varying effectiveness across multiple timescales. Network operators, however, favor a versatile predictor capable of performing well across diverse scales and prediction tasks.

In this study, we examine how these challenging characteristics affect deep forecasting models across different network environments. In the following section, we provide a formal definition of the network traffic prediction problem.

\subsection{The Network Traffic Prediction Problem}

A time series is a sequence of \(T\) observations \(\{x_1, x_2, \dots, x_T\} \in \mathbb{R}^{T \times d}\), where each \(x_t\in\mathbb{R}^d \) represents the observed values at time point \(t\), and \(d\) is the dimensionality of the data, i.e., the number of variables. These observations are ordered in time, with the time interval between successive observations denoted as \(\Delta t \) defining the timescale of observations.
Time series forecasting aims to model temporal dependencies in past and current observations to predict future values. 

Specifically, let \(x_t \in \mathbb{R}^d \) denote the multivariate input vector at time \(t \), which includes network traffic volume in bytes as well as exogenous variables such as weather conditions or holidays. 
The problem of network traffic prediction involves estimating the future sequence of traffic values \(\mathbf{Y}_{T+H}=\{y_{T+1}, y_{T+2}, \dots, y_{T+H}\}\in \mathbb{R}^{H \times 1}\), given a historical input window of length \( m \), denoted as \( \mathbf{X}_T = \{ x_{T-m}, x_{T-m+1}, \dots, x_T \} \in \mathbb{R}^{m \times d} \). Here, \( T \) represents the current time step, \( H \) is the prediction horizon, and \( d \) is the number of input features. The target variable \( y_t \) represents the traffic volume at time \(t\), which is a component of the input vector \( x_t \).
The network traffic prediction task can be formulated as a learning problem, where the goal is to estimate future traffic values based on the observed historical data. Formally, this is expressed as:

\begin{equation}
\label{eqn:forecasting_problem}
    \mathbf{\hat{Y}}_{T+H} = \mathcal{F}(\mathbf{X}_T),
\end{equation}

where \( \mathcal{F} \) is a predictive model that maps the input sequence \( \mathbf{X}_T \in \mathbb{R}^{m \times d} \) to the predicted future values \( \mathbf{\hat{Y}}_{T+H} \in \mathbb{R}^{H \times 1} \). The objective is to minimize an error metric \( \mathcal{L}(\mathbf{\hat{Y}}_{T+H}, \mathbf{Y}_{T+H}) \), which quantifies the difference between the predicted sequence and the ground truth.
Predictions are classified as short-term or long-term based on the forecast horizon \(H\). This study examines short-term predictions over one to several hours, essential for optimization and anomaly detection, and long-term predictions spanning days or months, relevant to capacity planning and energy conservation. 
The subsequent section reviews tradition statistical methods for network traffic prediction.

\subsection{Traditional Network Traffic Forecasting Techniques}
Naive methods that reuse past values and statistical models such as ARIMA and Exponential Smoothing, which leverage mathematical techniques to detect patterns and handle stochastic variation, serve as baseline benchmarks for network traffic prediction, offering simple and interpretable standards for assessing more advanced models.

\subsubsection{Naïve Forecasts}
The Naïve forecasting method predicts future values by repeating the most recent observation. Alternatively, for periodic data, a seasonal Naïve forecast assumes future values will match those observed in the same period of the previous cycle as \(\mathbf{\hat{Y}}_{t+H} = \mathbf{\hat{Y}}_{t+H-k}\) where \(k\) is the length of the seasonal cycle set to one week \cite{jiang2022internet}, as it is the largest observed cycle for traffic data. While simple and effective for highly regular data, this method performs poorly in the presence of sudden changes or non-seasonal fluctuations.

\subsubsection{ARIMA Model} ARIMA combines past observations (AutoRegressive term) and past forecast errors (Moving Average terms) to model a time series. The ARIMA\(p,d,q)\) model is given by 
\(
    \phi_p(L)(1-L)^d x_t = \theta_q(L)e_t
\)\cite{box1976analysis}
where \(x_t\) denote the time series; \(e_t\) is the error term; \(L\) is the lag operator 
\(phi_p(L)\) and \(\theta_q(L)e_t\) are the AR and MA polynomials of order \(p\) and \(q\); \(d\) is the differencing order used to achieve stationarity.

A seasonal extension, SARIMA, incorporates periodic patterns and is denoted as \(ARIMA(p, d, q)(P_1, D_1, Q_1)\) defined by
\(
    \phi_p(L)\Phi_{P_1}(L^{K_1})(1-L)^d(1-L)^{D_1}x_t = \theta_q(L)\Theta_{Q_1}(L^{k_1})e_t
\)
where \(K_1\) is the seasonal period; \(\Phi_{P_1}\) and \(\Theta_{Q_1}\) are seasoanl AR and MA polynomial..
Model parameters are typically estimated statistically and selected using criteria such as AIC to balance goodness of fit and model complexity.

\subsubsection{Exponential Smoothing (ES)}
ES models base predictions on underlying patterns of trend and seasonality, while smoothing out random noise by through weighted averages of historical values. The models is defined as follows\cite{wheelwright1998forecasting}.
\begin{align}
    \label{eqn:exopentialsmoothing}
    S_t &= \alpha\frac{x_t}{D_{t - k_1}} + (1-\alpha)(S_{t-1}+T_{t-1})\\
    T_t &= \beta (S_t - S_{t-1}) + (1-\beta)T_{t-1} \\
    D_t &= \gamma\frac{x_t}{S_t} + (1-\gamma)D_{t-K_1} \\
    \hat{y}_{t+h} &= (S_t + hT_t) \times D_{t-K_1+h}
\end{align}

where \(S_t\), \(T_t\), and \(D_t\) stand for the level, trend and seasonal estimates, \(K_1\) is the seasonal period, and \(\alpha\), \(\beta\), and \(\gamma\) are smoothing parameters. The parameters are selected via a grid search to minimize the training error.

Statistical methods effectively predict network traffic when temporal dynamics are simple, especially at minute scales. However, due to the high dimensionality and non-stationarity of network traffic time series, research has shifted toward deep learning models. The subsequent section reviews related work in this area.

\section{Related Work and Motivation}
\label{sec:related_work}

This section begins with an overview of pure deep forecasting models used in network traffic prediction, discussed in Section 
\ref{sub_related_work:NTPmodels}. We then evaluate comparative studies of these models in Section 
\ref{sub_related_work:NTPBenchmarks}. Finally, we identify gaps in existing research, considering recent developments and motivation in the time series forecasting field, and highlight our contributions to addressing these emerging questions in Section 
\ref{sub_related_work:TSFFindings&Motivation}.

\subsection{Deep Forecasting Models in Network Traffic Prediction}
\label{sub_related_work:NTPmodels}
Deep forecasting models have attracted significant attention within the NTP community due to their ability to learn non-linear relationships from large datasets, aiding network traffic prediction across various temporal scales and forecast horizons. 

The Multi-Layer Perceptron (MLP), considered the simplest form of deep learning architecture, was used by Cortez et al.\cite{cortez2006internet} to develop an ensemble of neural networks to predict TCP/IP traffic at five-minute and hourly intervals. Their model demonstrated superior accuracy compared to ARIMA and Exponential Smoothing models and achieved reduced training and testing durations, early indicating its potential effectiveness for network traffic prediction.
Meanwhile, Miguel et al.\cite{miguel2012new} proposed a methodology for aggregating traffic volume data between Autonomous Systems (AS) into time series formats using the NetFlow protocol. They designed an ensemble of Time Lagged Feed-Forward Networks (TLFNs) that incorporated short-term memory within the input layer to effectively manage temporal data. Their approach showed competitive performance relative to exponential smoothing methods and proved suitable for traffic routing applications. 

Recurrent Neural Network (RNN) based models, inherently designed for sequential data, have been increasingly popular for various NTP tasks. Romakrishnan et al. \cite{ramakrishnan2018network} collected two hours of packet traces using Wireshark on virtual machines and utilized two backbone traffic datasets. They evaluated the performance of RNNs, Long Short-Term Memory networks (LSTMs), and Gated Recurrent Units (GRUs) in predicting traffic volume, packet protocols, and packet distribution. Their results demonstrated that LSTM and GRU models outperformed the ARIMA model by up to 78\% on the collected dataset, particularly in predicting traffic spikes and fluctuations. Additionally, Azzouni et al\cite{azzouni2018neutm} proposed a framework called neuTM that leverages LSTM to learn from historical traffic data and predict future network traffic metrics. The LSTM model was evaluated for single-point prediction at 15-minute intervals and significantly outperformed statistical models. 

Additionally, Fischer et al.\cite{fischer2020deepflow} introduced DEEPFLOW, a traffic prediction system designed to analyze ingress traffic data. This system employs a combination of statistical models and deep learning techniques, including various sequence modeling architectures such as LSTM and Sequence-to-Sequence LSTM, to generate comprehensive traffic flow predictions.
Another work that utilized sequence-to-sequence models is by Feng et al.\cite{feng2018deeptp}, which proposed DeepTP, a model designed to predict traffic demand while accounting for temporal variations and external factors. The model consists of a general feature extractor that encodes external information, and a sequential module that captures temporal dynamics. The authors employed a sequence-to-sequence architecture with an attention mechanism to directly predict multiple future steps, thereby enhancing model performance. Results demonstrate that incorporating external information significantly improves traffic prediction accuracy in real scenarios.

To model traffic nonlinearities, Bi et al.\cite{bi2021hybrid} developed a two-step architecture that combines a Temporal Convolution Network (TCN) and LSTM. After filtering the traffic time series to remove strong noise, TCN extracts short-term local features, followed by an LSTM that captures long-term dependencies and generates predictions. Their hybrid method significantly outperformed statistical baselines and individual TCN and LSTM models. 

More recently, the Transformer architecture, which has revolutionized sequence modeling, has also been adopted in this context.
Kougioumtzidis et al.\cite{kougioumtzidis2025mobile} employed the Temporal Fusion Transformer (TFT) for traffic volume forecasting. Their attention-based model effectively captured complex dependencies within traffic time series data and distinguished between weekend, weekday, and holiday patterns in 24-hour ahead predictions. The evaluation results indicated that the TFT model outperformed the statistical baseline, including TCN and LSTM models. However, the authors also noted that the computational complexity of the TFT model is substantially higher than that of these baseline models. 

Additionally, Pérez et al. \cite{10620742} evaluated the performance of two advanced forecasting models, DLinear and PatchTST, in comparison to the Long Short-Term Memory (LSTM) model. The evaluation was conducted using two traffic-volume datasets, each recorded at 10-minute intervals. The results indicated that both LSTM and PatchTST outperformed DLinear, with LSTM surpassing both models on one dataset. Additionally, all models demonstrated superior accuracy in predicting abnormal patterns, such as weekends, on which DLinear struggled to forecast weekend-specific patterns but maintained satisfactory performance otherwise.

\textbf{The Gap.} Despite the promising potential of advanced TSF models, their adoption within the NTP community remains limited due to several major challenges: the scarcity of readily available implementations, the complexity and variability of hyperparameters, and the perceived high computational costs.
Our study identify eight promising TSF models covering diverse architectures, discuss thier innovative processing capabilities and emprically test their performance, robustenss, and effectiveness over broad NTP tasks.

\begin{table*}
    \caption{Comparison of Our Contributions with Existing Work}
    \label{tab:contribution}
    \centering
    \begin{tabular}{>{\raggedright\arraybackslash}p{0.18\textwidth} 
                    >{\raggedright\arraybackslash}p{0.36\textwidth} 
                    >{\raggedright\arraybackslash}p{0.35\textwidth}}
    \toprule
    \textbf{Contribution} & \textbf{Our Work} & \textbf{Existing Work} \\
    \midrule
    \textbf{Identifying, Implementing, and Evaluating Advanced TSF Models for NTP} & 
    
    We identify promising TSF trends and implement and evaluate 12 deep forecasting models, each with unique advantages that address the challenges of network traffic prediction for multiscale and multi-horizon NTP. &
    Despite recent advances in the TSF field with diverse architectures and complexities, comparative studies on network traffic prediction mainly focus on RNN architectures evaluated on a single timescale. \\
    \midrule
    \textbf{Developing a Systematic Methodology and Comprehensive NTP Benchmark} &  We analyze the impact of traffic challenges—such as anomalies, missing data, and externally influenced variations—on \(4\) distinct network traffic datasets, reflecting real-world challenges. &
    To our knowledge, there has been no comprehensive assessment of model performance on real-world traffic characteristics across various network environments and timescales in current traffic prediction research. \\
    \midrule
    \textbf{Model Behavior Insights and Selection Guidelines}
    & We connect model behavior to traffic characteristics and deployment constraints, identifying performance regimes and efficiency thresholds that inform model selection and future development.
    & Existing work primarily compares predictive performance, with limited attention to data efficiency, tuning effort, and energy consumption or their implications for model selection. \\
    \bottomrule
    \end{tabular}
    \label{tab:contribution}
\end{table*}

\subsection{Comparative Evaluations of Network Traffic Predictors}
\label{sub_related_work:NTPBenchmarks}
There have been considerable research efforts to empirically compare the performance of deep forecasting models on network traffic prediction.
Oliveira et al.\cite{oliveira2016computer} evaluated two deep learning architectures, MLPs and RNNs, for predicting traffic traces at 5-minute, hourly, and daily intervals. Their results indicated that RNNs achieved the lowest prediction error and required fewer neurons, making training up to 66\% faster than MLPs with over 8\% higher accuracy.
Jian et al.\cite{jiang2022internet} compared thirteen deep learning models against five baselines using a \(6\)-month open Internet bandwidth utilization dataset. The results showed that all deep learning models outperformed the baselines, with InceptionTime—a CNN ensemble—achieving the lowest error. 

In recent studies, Ferreira et al. \cite{ferreira2023forecasting} conducted experiments with Simple RNNs, LSTMs, and GRUs, comparing their performance with SARIMA baselines for hourly traffic prediction and analyzing the effects of layer and neuron counts on accuracy. Their results indicated that at least one configuration of each deep learning model can outperform SARIMA. Conversely, Saha et al. \cite{saha2023outliers} examined the impact of traffic outliers on multistep traffic prediction using deep sequence models, including RNNs, LSTMs, GRUs, and sequence-to-sequence LSTMs with and without attention. They used a 29-day ISP traffic dataset sampled every 5 minutes to evaluate models with and without anomalies for predictions up to 12 steps (1 hour). Their findings showed that the LSTM encoder-decoder performs best even with anomalies, and all models improve after outlier removal.

Aouedi et al.\cite{aouedi2025deep} compared the performance of five deep learning models—LSTM, BiLSTM, GRU, BiGRU, and GWN—across three backbone traffic datasets. They assessed prediction error, inference time, and sequence length. All models improved performance as sequence length increased, indicating better retention of early time steps. The GWN model achieved the highest accuracy but required longer inference times.
Koumar et al. \cite{koumar2025comparative} evaluated seven deep learning models on a large ISP traffic dataset spanning approximately one year. The models included five RNNs (GRU, LSTM, GRU-FCN, LSTM-FCN, RCLSTM) and two advanced CNNs (InceptionTime, ResNet). They assessed hourly traffic prediction across various traffic series granularities, and from one hour to one week ahead. Missing data were imputed with zeros, as it indicated an inactive period on smaller granularities, such as on IP series. Results showed a negative correlation between the proportion of missing data and prediction accuracy. Classical architectures like LSTM and GRU performed similarly to their convolutional variants and outperformed ResNet and InceptionTime. The GRU-FCN model achieved the fastest training and prediction times. 

\textbf{The Gap.}
Existing comparative studies summarized in Table~\ref{tab:related_work} primarily focus on RNNs and CNNs for short or single-step predictions over limited timescales, providing a narrow basis for model selection. Some consider traffic granularity, missing data, sequence length, or computational cost, but do not systematically relate model behavior to traffic characteristics and deployment conditions. Additonally, efficiency assessments largely emphasize training and testing time, with limited attention to data efficiency, tuning effort, and energy consumption.
In contrast, our study jointly evaluates model accuracy, robustness, data efficiency, tuning requirements, and resource efficiency across multiple network environments and timescales. More importantly, we synthesize these findings to characterize how different forecasting architectures behave under changing traffic and deployment conditions, and translate these behaviors into practical guidelines for model selection across network-management tasks.

\begin{figure*}[t!]
    \centering
    \includegraphics[width=0.95\textwidth]{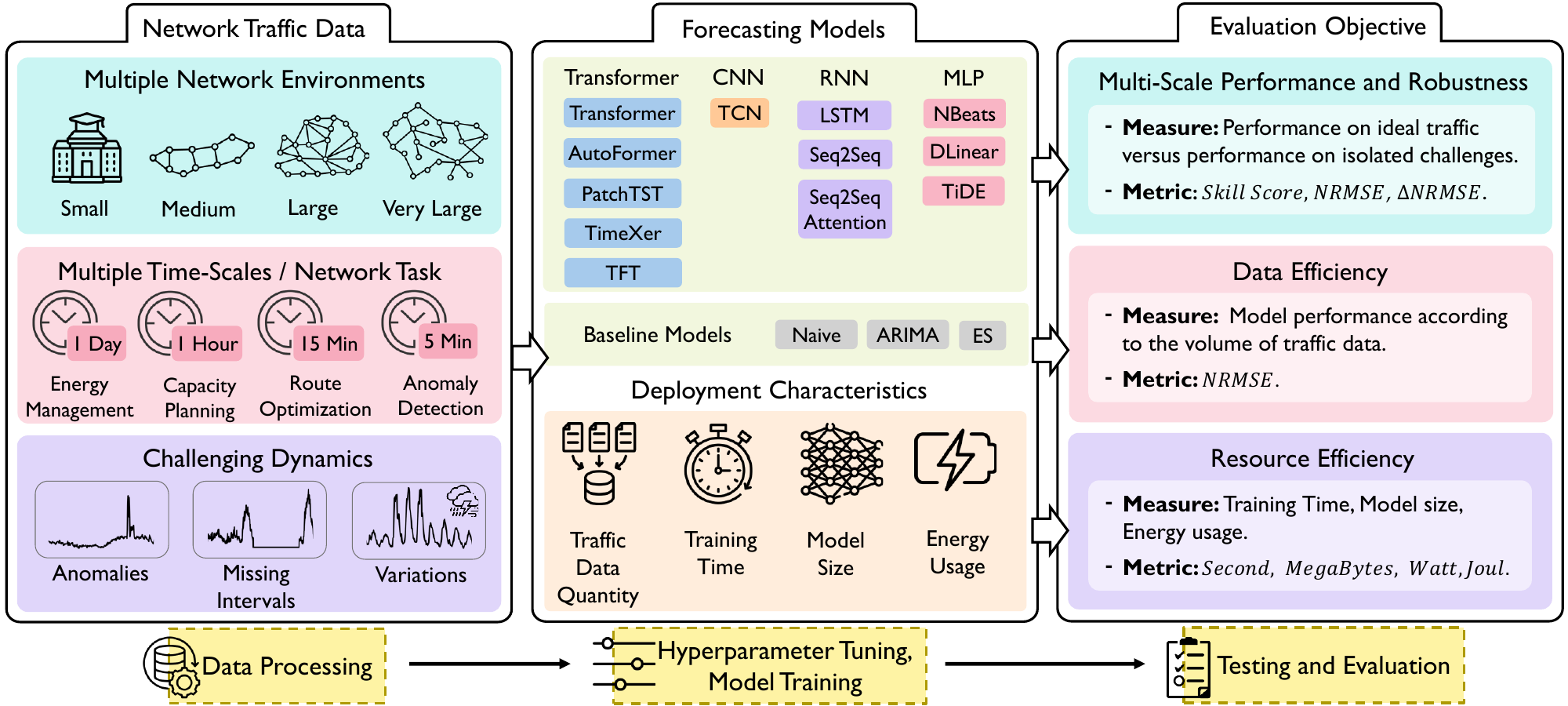}
    \caption{
    Our proposed evaluation framework. The upper part shows core components, and the bottom strip illustrates the unified training and evaluation process.
    }
    \label{fig:methodology}
\end{figure*}

\subsection{Advances and Findings in Time Series Forecasting}
\label{sub_related_work:TSFFindings&Motivation}
The network traffic prediction problem is a special class within time series forecasting, and as such, it is influenced by the advancements and challenges inherent to this field.

Makridakis et al.\cite{makridakis2022m5} examined the findings of the M5 time series forecasting competition, a high profile forecasting challenge with 30,490 time series and approximately 90,000 model submissions for the accuracy track. They report that the vast majority
 (about 92.5\%) failed to outperform a simple Exponential Smoothing (ES)
 benchmark, indicating that the use of ML and DL methods does not guarantee superior performance. The study also highlights the importance of incorporating exogenous variables to enhance forecasting accuracy, as demonstrated by the top-performing models.

Additionally, Zeng et al.\cite{transformer_effective_2023} questioned the validity of transformer-based models for long-term time series forecasting. Their study showed that the performance gains of advanced transformer models are primarily due to their direct multistep prediction capability rather than their architectures. To corroborate this, they developed simple one-layer models, LSTF-Linear and DLinear (linear layer with decomposition), and compared their performance against four transformer-based models—FEDformer, Autoformer, Informer, and Pyraformer—across nine datasets. The results indicated that these simple linear models can outperform more sophisticated models, with time and memory requirements that are orders of magnitude lower. 

However, a recent time series benchmarking study challenges this view and confirms the superiority of transformer models. Wang et al.\cite{wang2024deep} evaluated advanced TSF models across diverse time series analysis tasks, including both short- and long-term forecasting. They demonstrated that transformer models exhibit highly competitive performance across all tasks, attributed to their inherently powerful data modeling capabilities. Additionally, they showed that advanced MLP models also perform competitively in forecasting.

\textbf{The Gap.}
These findings challenge the assumed superiority of DL models over statistical baselines and question whether advanced TSF models are necessary beyond simpler DL architectures. This raises important questions for deployment choices and model exploration in network traffic prediction concerning the trade-offs between accuracy and efficiency. Our study addresses this gap by systematically comparing DL models and advanced TSF models against robust statistical baselines. We analyze how key challenging features of traffic data impact model accuracy and assess data and resource efficiency, highlighting models' strengths and limitations. Our work offers practical insights for both researchers developing new prediction models and practitioners choosing suitable solutions for real-world NTP tasks. Our main contributions, compared to existing work, are summarized in Table \ref{tab:contribution}.

\section{Evaluation Framework of Deep Network Traffic Forecasting Models}
\label{sec:methodology}

This section presents our network traffic datasets and analyzes their characteristics in \ref{subsec:datasets}. We then examine promising TSF trends using selected deep forecasting models and discuss their innovations in \ref{subsec:evaluated_models}. Lastly, we outline the experimental setup, objectives, and evaluation metrics in \ref{subsec:experiments}. Our overall framework is illustrated in Fig. \ref{fig:methodology}.

\subsection{Network Traffic Datasets}
\label{subsec:datasets}
\subsubsection*{Temporal Dynamics}
\begin{figure*}
    \centering
    \includegraphics[width=\linewidth]{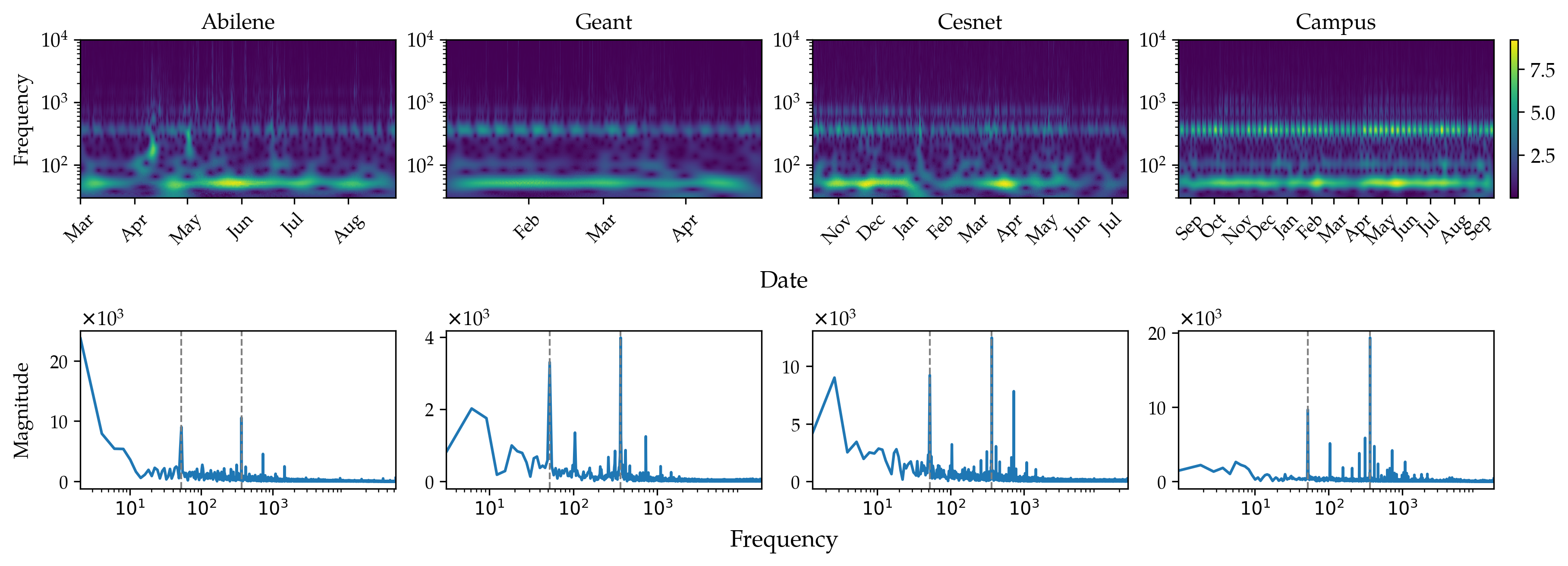}
    \caption{CWT Spectrograms and Corresponding FFT Plots for the \(4\) Traffic Datasets. Each dataset is represented by two figures: one for the CWT spectrogram and one for the corresponding FFT plot. The bottom row presents the FFT plots, with marked peaks indicating the dominant daily and weekly frequencies. The top row shows the corresponding CWT spectrograms, highlighting the time-frequency distribution of these frequencies. }
    \label{fig:frequency}
\end{figure*}

We use four real-world traffic datasets that cover a range of network environments. This variety helps capture the complex dynamics of network traffic in different contexts and supports the generalization of our results. 

\subsubsection{Abilene \cite{abilene}}
The Abilene dataset was collected from a backbone network in North America, comprising \(12\) interconnected router nodes. It records moderate traffic volumes, aggregated over \(5\)-minute intervals for origin-destination flows and spans six months, from March 1 to September 10, 2004.

\subsubsection{GEANT \cite{uhlig2006geant}}
The GÉANT network, which also includes traffic metrics similar to those of Abilene, represents a larger and more diverse backbone infrastructure. It consists of 23 nodes and 36 links, with data aggregated every \(15\) minutes over a four-month period beginning January 8, 2005. 

\subsubsection{CESNET \cite{koumar2025cesnet}}
The CESNET dataset is a relatively recent collection encompassing approximately ten months, from October 2023 to July 2024. It represents a large-scale modern ISP backbone network serving hundreds of institutions across the Czech Republic. The dataset, aggregated at \(10\) minutes scale, includes substantial traffic volume, comprising billions of IP flows, and primarily captures real-world ISP traffic characteristics. The dataset is supplemented with calendar information for public holidays.

\subsubsection{Campus Network}
This dataset comprises recent traffic data collected from our campus network, which supports thousands of connected hosts, over nearly a year, from August 16, 2024, to September 19, 2025. Traffic flows are captured from the internet-facing router using the NetFlow protocol and aggregated at \(5\) minute intervals. This dataset provides a detailed view of traffic dynamics, highlighting pronounced multiseasonal patterns and variations due to external factors.

We analyze the periodicities and temporal dynamics of each dataset in the frequency domain using the Discrete Fourier Transform (DFT).
\begin{equation}
    X_k = \sum_{n=0}^{N-1} x_n e^{-j \frac{2\pi k n}{N}}
\end{equation}

where \( X_k \) represents the \( k \)-th frequency component for \( k = 0, 1, \dots, N-1 \), \( x_n \) is the \( n \)-th sample in the input time-domain sequence \( \{x_t\}_{t=1}^T \), and \( N \) is the total number of samples.  
The power spectral density (PSD) indicates the magnitude of each frequency component in the signal. As shown in the bottom row of Fig. \ref{fig:frequency}, all datasets display two primary peaks at frequencies corresponding to daily (\(365\)) and weekly (\(52\)) cycles, with smaller peaks at half a day (\(730\)).

We further analyze how these frequency components evolve over time via Continuous Wavelet Transform (CWT), which provides a time-frequency representation as follows.

\begin{equation}
    \text{CWT}(s, t) = \int_{-\infty}^{\infty} x(\tau) \, \psi^* \left( \frac{\tau - t}{s} \right) d\tau
\end{equation}

where \( x(\tau) \) is the signal, \( \psi \) is the wavelet function, \( s \) is the scale (related to frequency), and \( t \) is the time shift. Unlike the DFT, which captures only global frequency information, CWT enables localized analysis of non-stationary signals like network traffic time series. 

As shown in the top row of Fig. \ref{fig:frequency}, the pulses represent daily frequency, interrupted by weekend patterns and holidays, while the continuous line at the bottom indicates weekly frequency. The weakened weekly frequency observed in April for Abilene and March for Geant results from missing values. Both the campus and Cesnet datasets also exhibit reduced frequency in January, linked to the year-end holiday period that lowers weekday traffic and disrupts regular patterns. A similar, though less pronounced, effect is observed for Geant in April; however, the reason for the decrease in traffic remains uncertain due to the dataset's age. Corresponding with the beginning of the new school year in April, campus traffic shows increased weekly, daily, and half-daily frequencies. The characteristics of the four datasets (section 
\ref{sub_background:traffic_data_characteristic}) and their derived time scale versions are quantified in Table 
\ref{tab:datasets}.

\begin{table}[]
\caption{Summary of Datasets and Their Native Characteristics}
\renewcommand{\arraystretch}{1.2}  

\noindent
\begin{tabularx}{\linewidth}{
p{0.01\columnwidth}
 >{\raggedright\arraybackslash}X
 >{\centering\arraybackslash}X 
 >{\centering\arraybackslash}X 
 >{\centering\arraybackslash}X 
 >{\centering\arraybackslash}X 
 >{\centering\arraybackslash}X}
\toprule
 & \textbf{Dataset Code} & \textbf{Scale (Minute)} & \textbf{Size (Points)} & \textbf{Missing (\%)} & \textbf{Outlier (\%)} & \textbf{Varient Days(\%)} \\
\midrule

\multirow{3}{*}{\rotatebox{90}{Abilene}} 
& A5M  & 5  & 55008  & 13.6 & 1.9   & -- \\
& A15M  & 15  & 18336  & 13.6 & 1.9   & -- \\
& A12H  & 720  & 382  & 13.6 & 1.9   & -- \\
\midrule

\multirow{2}{*}{\rotatebox{90}{Geant}} 
&G15M & 15 & 11394  & 5.5  & 0.21  & -- \\
&     &   &   &   &   &   \\
\midrule

\multirow{3}{*}{\rotatebox{90}{Cesnet}} 
&C10M & 10 & 40308  & 0.6  & 0.31  & -- \\
&C1H &  60 & 6718  & 0.6  & 0.31  & 6.7 \\
&C1D &  1440 & 280  & 0.6  & 0.31  & 6.7 \\
\midrule

\multirow{4}{*}{\rotatebox{90}{Campus}} 
&K5M  & 5  & 114912 & 7.3  & 0.005 & -- \\
&K15M & 15  & 38305 & 7.3  & 0.005 & -- \\
&K1H  & 60  & 9576 & 7.3  & 0.005 & 6.2 \\
&K12H & 720  & 800 & 7.3  & 0.005 & 6.2 \\
\bottomrule

\end{tabularx}
\label{tab:datasets}
\end{table}

\subsection{Deep Network Traffic Forecasting Models}
\label{subsec:evaluated_models}

We identify \(8\) advanced TSF models, each pioneering a promising trend, and \(4\) traditional deep learning models, each offering unique advantages for network traffic prediction based on data characteristics and forecasting horizon.

\subsubsection{Long Short-Term Memory (LSTM) \(1997\)}
The LSTM, developed through \cite{hochreiter1997long, gers2000learning}, is a fundamental sequence model. It employs \textit{gates} to regulate information flow and mitigate the vanishing gradient problem inherent in simple RNNs. Through \textit{forget, input,} and \textit{output} gates, LSTM retains and manages information over long sequences, effectively capturing complex dependencies and nonlinear patterns. 

\subsubsection{Sequence to Sequence (Seq2Seq) \(2014\)}
The Seq2Seq model \cite{sutskever2014sequence} utilizes two LSTMs: an encoder and a decoder. The encoder processes input sequences to generate a final hidden state, the \textit{context vector}, summarizing temporal information. The decoder initializes with this vector, then sequentially produces the output sequence. This architecture effectively handles variable-length predictions by enabling complex nonlinear transformations of the input. 

\subsubsection{Sequence to Sequence with Attention (Seq2Seq) \(2014\)}
The \textit{attention} mechanism\cite{bahdanau2014neural} overcomes the bottleneck in standard Seq2Seq models by supplying the decoder with a dynamic summary of the input at each output step, rather than relying on a single context vector. This direct link to all encoder hidden states improves the modeling of long-term dependencies, albeit with increased computational complexity.

\subsubsection{Transformer \(2017\)}
The Transformer, introduced in \cite{vaswani2017attention}, is a scalable encoder-decoder architecture that revolutionized sequence modeling by eliminating recurrence and relying solely on attention mechanisms and feedforward networks for processing sequential data. Self-attention evaluates the importance of each time step, while multiple attention heads enable parallel processing to capture various dependencies. 


\subsubsection{Autoformer \(2021\)}
Autoformer \cite{Autoformer2021} replaces canonical self-attention with an Auto-Correlation mechanism that discovers period-based dependencies and aggregates similar sub-series using Fast Fourier Transform for efficient computation. Decomposition blocks are embedded throughout the encoder and decoder to progressively separate trend and seasonal components, with the decoder accumulating the trend component while refining the seasonal representation. This design targets long-term forecasting while reducing the quadratic complexity of point-wise attention.


\subsubsection{Patch Time-Series Transformer (PatchTST) \(2022\)}
PatchTST \cite{nie2022time} adapts the Transformer by dividing each univariate channel into fixed-length subseries patches (overlapping or non-overlapping) that serve as input tokens to a shared Transformer encoder. The encoded patch representations are then mapped to the prediction horizon through a forecasting head. Additionally, PatchTST's channel-independent design applies attention across patches rather than individual time points, preserving local information while reducing the effective sequence length and enabling longer look-back windows.

\subsubsection{TimeXer \(2024\)}
TimeXer \cite{wang2024timexer} extends the transformer to explicitly incorporate exogenous variables. It uses patch-level representations for endogenous (target) data and sequence-level representations for exogenous data. A single global endogenous token summarizes the temporal information of the endogenous series and links it to the exogenous series, capturing their correlations. This approach enhances robustness in real-world applications by effectively targeting external factors.


\subsubsection{Temporal Fusion Transformer (TFT) \(2021\)}
The TFT \cite{TFT2021}  combines recurrent sequence modeling, attention, variable selection, and gating for multi-horizon forecasting with heterogeneous inputs. Variable selection networks filter relevant static, historical, and known future inputs, while recurrent encoder-decoder layers model local temporal patterns, and self-attention captures longer-range dependencies. Gated residual components regulate information flow and suppress unnecessary transformations. TFT effectively models nonlinear interactions among variables.

\subsubsection{Temporal Convolutional Networks (TCN) \(2018\)}
TCNs \cite{TCN2018} are convolutional neural networks (CNNs) tailored for sequence modeling. They utilize causal convolution to prevent future data leakage and dilated convolution to capture long-range dependencies. The dilation factor enables the network to skip inputs, thereby enlarging the receptive field without increasing the number of parameters. TCNs provide faster training compared to RNNs and effectively model complex nonlinear patterns in time series data.

\subsubsection{Decomposition Linear Model (DLinear) \(2023\)}
DLinear, introduced in \cite{transformer_effective_2023}, is a simple linear model that challenges the dominance of transformer architectures. It employs a series-decomposition approach similar to Autoformer, applying a moving-average filter followed by independent linear mappings of the decomposed components via a single-layer model. DLinear is efficient and effectively captures seasonal and macro-level temporal patterns.


\subsubsection{Neural Basis Expansion Analysis for Time Series (NBeats) \(2019\)}
NBeats \cite{Nbeats2019} is a deep architecture of stacked fully connected blocks that uses basis expansion to extract key components of a time series. Each block produces a backcast that is removed from its input and a forecast that contributes to the final prediction, enabling deep residual stacking. Its blocks use basis expansions that can be generic or structured to represent trend and seasonality, providing an implicit additive decomposition of the series that captures nonlinear and complex short-term relationships.


\subsubsection{Time-series Dense Encoder (TiDE) \(2023\)}
TiDE \cite{das2023long} is an MLP-based encoder-decoder architecture designed to incorporate historical and exogenous features. Dynamic covariates are first projected to a lower-dimensional representation and combined with the look-back sequence in a dense encoder; a dense decoder maps this encoding across the forecast horizon, while a temporal decoder reintroduces future covariates at each prediction step. Residual MLP blocks and a global linear residual connection provide nonlinear modeling while retaining a direct linear path from the historical input to the forecast. TiDE's deep MLPs effectively model complex nonlinear interactions between historical data and exogenous variables, achieving transformer-level performance at lower computational cost.

\begin{table*}
\caption{Deep Forecasting Models Summary. Advanced TSF models are bolded, and support for exogenous variables is underlined.\label{tab:model_advantage}}
 
\centering
\renewcommand{\arraystretch}{1.2}  
\setlength{\tabcolsep}{5pt}  
\begin{tabularx}{\textwidth}{
l 
 p{0.17\linewidth} 
 >{\centering\arraybackslash}X 
 c 
 p{0.3\linewidth} 
 p{0.14\linewidth} }
\toprule
\textbf{Model} & \textbf{Architectural Paradigm} & \textbf{Computation Complexity} & \textbf{Processing} & \textbf{Innovative Time Series Processing Logic} & \makecell[l]{\textbf{Key Hyperparamters}} \\
\midrule
\underline{LSTM}\cite{hochreiter1997long,gers2000learning} & Sequential Recurrent & \(O(L)\) & Sequential & Uses gated memory cells to retain long-term dependencies in sequential data. & Hidden Size, Number of Layers. \\

\underline{Seq2Seq}\cite{sutskever2014sequence} & RNN Encoder--Decoder & \(O(L)\) & Sequential & Encodes input sequence into a context vector for a decoder to generate variable-length outputs. & Hidden size, encoder layers, decoder layers. \\

\underline{Seq2Seq\_Attn}\cite{bahdanau2014neural} & RNN Encoder--Decoder with Attention &  \(O(L\cdot L\prime)\) & Sequential & Introduced Attention to dynamically access the entire input history at each decoding step. & Hidden size, encoder layers, decoder layers. \\

\midrule
\textbf{N-BEATS}\cite{Nbeats2019} & Stack of MLP Blocks & \(O(1)\) & Parallel & Stacked blocks that learn basis expansions for implicit trend-seasonality decomposition. & Block layers, layer size, block types. \\

\textbf{DLinear}\cite{transformer_effective_2023} & Decomposition with Linear Layer &  \(O(1)\) & Parallel & Decomposes series into trend and seasonal parts, each forecasted via a simple linear layer. & Moving average size \\
\underline{\textbf{TiDE}}\cite{das2023long} & MLP Encoder--Decoder & \(O(1)\) & Parallel & Uses residual MLPs to transform historical and exogenous inputs into a hidden state projected over the future horizon. & Hidden size, encoder layers, decoder layers. \\
\midrule
TCN\cite{TCN2018} & Dilated Convolution & \(O(L)\) & Parallel & Exponential expands receptive field with dilated convolutions to capture long-range dependecies. & kernal size, layers, Dilation rate \\

\midrule
\textbf{Transformer}\cite{vaswani2017attention} & Attention based Encoder--Decoder & \(O(L^2)\) & Parallel & Captures global long-term and shor-term dependencies using multi-head and self-attention without recurrence. & Attention heads, feed forward and model dimension. \\
\textbf{Autoformer}\cite{Autoformer2021} & Attention Encoder--Decoder &  \(O(LlogL)\) & Parallel & Explicitly decomposes series and replaces self-attention with an auto-correlation mechanism. &  Moving average size, attention factor. \\
\textbf{PatchTST}\cite{nie2022time} & Attention Encoder &  \(O((L/P)^2)\) & Parallel & Applies attention on sequence patches for extended context and local feature preservation. & Patch length, attention heads, layers. \\

\underline{\textbf{TimeXer}}\cite{wang2024timexer} & Transformer Encoder--Decoder &  \(O((L/P)^2)\) & Parallel & Fuses exogenous and target series via hierarchical representation for enhanced modeling. & Patch length, endcoder layers, decoder layers.  \\
\midrule
\underline{\textbf{TFT}}\cite{TFT2021} & Gated Feature Attention & \(O(L^2)\) & \multirow{2}{*}{\makecell{Sequential/\\Parallel}} & Combines gating, recurrence, and variable selection networks for multi-feature forecasting. & Hidden size, attention heads, encoder layers, decoder layers.\\
\bottomrule
\multicolumn{6}{l}{\makecell[l]{ \(L\) stands for input sequence lengths, while \(L\prime\) stands for output sequency length.  }}
\end{tabularx}
\end{table*}

\subsubsection*{Models Implementation Details}
These relatively recent models are not yet standardized or readily available as modules in mainstream deep-learning libraries. Their official implementations are provided exclusively as open-source code in the authors’ GitHub repositories accompanying their publications. The recently released TSLib \cite{wang2024deep} consolidates many contemporary models, but it did not include all the architectures examined here at the time of this study. To ensure completeness and consistency, we adapted and integrated the LSTM, Seq2Seq, Seq2Seq\_Attn, TCN, and NBeats. All models in this study are implemented in \textit{PyTorch} for its flexibility in integrating computational measurements.
A summary of the compared models-including their computational complexity, innovations, and key hyperparameters-is presented in Table \ref{tab:model_advantage}

\subsection{Experimental scenarios}
\label{subsec:experiments}

At a high level, our experiments are designed to address the following key RQs, which we devise to inform the development of deep forecasting models for network traffic prediction.

\textbf{RQ1} Multi-Scale consistency and robustness: Which forecasting model exhibits the strongest generalization capabilities across NTP tasks, while demonstrating robustness to challenging traffic characteristics?

\textbf{RQ2} Data Efficiency:  How do various network traffic forecasting models compare in terms of data efficiency, particularly in maintaining accuracy when trained with limited data?

\textbf{RQ3} Resource Efficiency: Which forecasting models offer the highest resource efficiency for deployment in low-resource network environments?

\subsubsection{NTP Tasks}
We define the forecasting tasks listed in Table 
\ref{tab:ntp_tasks} for our comparative evaluations. All tasks use Direct Multi-Step (DMS) forecasting, which is more practical for real-world network traffic prediction. Advanced TSF models, particularly encoder-decoder architectures, can directly generate variable-length output sequences, reducing error accumulation compared to recursive single-step forecasting. 

We set input lengths to capture traffic periodicities of one week, one day, and half a day, based on FFT analysis shown in Fig. 
\ref{fig:frequency}. Due to the large size of \(5\)-minute datasets, the input length was limited to \(96\) points (a third of a day) to enhance training efficiency. All datasets exhibit frequency peaks at eight hours. Prediction horizons were selected to align with typical NTP tasks: hours ahead for anomaly detection, daily forecasts for energy saving and capacity planning, and monthly horizons for long-term planning, with a one-month horizon used due to limited data for larger aggregations. 

\begin{table}[t!]
\renewcommand{\arraystretch}{1.2}  
\setlength{\tabcolsep}{5pt}        
\centering
\caption{NTP Tasks, time scale, and prediction horizons across datasets}
\label{tab:scale_input_output_horizon}
\begin{tabularx}{\linewidth}{l 
c
>{\centering\arraybackslash}X 
>{\centering\arraybackslash}X 
>{\centering\arraybackslash}X 
}
\toprule
\textbf{Time Scale} & \textbf{Datasets} & \textbf{Input Length} & \textbf{Output Length} & \textbf{Prediction Horizon} \\
\midrule
5 minutes  & A5M, K5M   & \(96\) & \(12\) & 1 hour \\
10 minutes & C10M   &  \(144\) & \(36\) & 6 hours \\
15 minutes & \makecell[c]{G15M, A15M,\\ K15M} & \(96\)  & \(24\) & 6 hours \\
1 hours    & K1H, C1H   &  168 & 24  & 1 Day \\
12 hours   & A12H, K12H    & 14     & 60 & 1 month  \\
1 Day      & C1D   & 7  & 30 & 1 month \\
\bottomrule
\end{tabularx}
\label{tab:ntp_tasks}
\end{table}


\subsubsection{Model Hyperparameter Tuning}
We conduct model-specific hyperparameter optimization (HPO) for each network-traffic forecasting task using Optuna. For traditional architectures, the search ranges follow previous comparative studies\cite{koumar2025cesnet, lara2021experimental}, while for advanced TSF models, the ranges and default values are derived from their respective works. The model-specific ranges are summarized in Table\ref{tab:hyperparameters}. Due to dataset variability, learning rate, dropout, and batch size are additionally optimized with architecture-specific parameters.

Each optimization study uses expanding-window time-series cross-validation configured with up to five valid folds. To provide a common practical optimization protocol across architectures, we limit each model--dataset search to 50 trials or a one-hour study timeout, while allowing running trials to finish. Consequently, observed HPO time can exceed one hour. Models are trained for up to 100 epochs with early stopping. Because model-specific search spaces and per-configuration computational costs differ, we deliberately adopt this common stopping rule to compare the improvements achievable under a resource-constrained tuning protocol, reflecting network operators' limited capacity for task-specific optimization. This provides a consistent basis for evaluating practical tunability, while allowing model-specific computational costs to determine the extent of configuration exploration.

\subsubsection{Standarized Hyperparamters}
We also conducted an experiment using standardized hyperparameters across models to evaluate the relative performance of deep forecasting architectures and their stability. The default hyperparameters are highlighted in bold in Table \ref{tab:hyperparameters}.

\begin{table}[t!]
\centering
\caption{Model Hyperparameters and Tuning Ranges. Standardized setup in bold.}
\label{tab:hyperparameters}
\begin{tabularx}{\linewidth}{lXX}
\toprule
\multicolumn{3}{c}{\textbf{Unique Model Hyperparameters}} \\ 
\midrule
\textbf{Model} & \textbf{Hyperparameter} & \textbf{Range} \\ 
\noalign{\hrule height 1.2pt}
\addlinespace[3pt]
\multirow{2}{*}{\makecell[l]{LSTM, Seq2Seq, \\ Seq2Seq\_Attn}} 
 & Hidden Size & \{32, 64, \textbf{128}, 256\} \\
 & LSTM Layers & (\textbf{1}, 4) \\

\multirow[t]{3}{*}{TCN} 
 & TCN Layers & (1, \textbf{4}) \\
 & Kernel Size & \{\textbf{3}, 5, 6\} \\
 & Dilation Base & \textbf{2} \\

\multirow[t]{4}{*}{NBeats} 
 & Number of Blocks & (2, \textbf{4}) \\
 & Number of Layers & (2, \textbf{4}) \\
 & Layer Size & \{\textbf{256}, 2048\} \\
 & Block Type & \{['trend','seasonality'], \textbf{'generic'}\} \\

DLinear & Moving Average & \{5, 15, \textbf{25}, 35\} \\
TiDE & Hidden Size & \{64, \textbf{128}, 256\} \\
PatchTST, TimeXer & Patch Length & \{8, \textbf{16}, 32\} \\
Autoformer  & Attention factor &  (1, 2, \textbf{3}, 4, 5) \\
TFT & Hidden Size & \{64, \textbf{128}, 256\} \\
\midrule

\multicolumn{3}{c}{\textbf{Encoder–Decoder Common Hyperparameters}} \\ 
\midrule
 & Encoder Layers & (1, \textbf{2}, 3) \\
 & Decoder Layers & (\textbf{1}, 2) \\

\midrule
\multicolumn{3}{c}{\textbf{Transformer Based Common Model Hyperparameters}} \\ 
\midrule
 & Attention Heads & \{2, 4, \textbf{8}\} \\
 & Model Dimension & \{256, \textbf{512}\} \\
 & Dimension Feed Forward & \{1024, \textbf{2048}\} \\
\midrule

\multicolumn{3}{c}{\textbf{General Model Hyperparameters}} \\ 
\midrule
 & Learning Rate & \{\textbf{1e2}, 1e5\} \\
 & Dropout & \{\textbf{0.1}, 0.2\} \\
 & Batchsize & \{16, \textbf{32}, 64\} \\
\bottomrule
\end{tabularx}
\end{table}

\subsubsection{Measurment Environment}
We conduct our experiments on a Desktop Computer with an Intel i5-12400,
CPU@4.4GHz, 32GB memory, NVIDIA RTX 3050, and a 64-bit Ubuntu 22 operating system.

\subsubsection{Evaluation Protocols}
We conduct three experiments, each targeting a specific research question (RQ). First, we assess the performance of all 12 deep forecasting models against naive and statistical models based on the tasks outlined in Table 
\ref{tab:ntp_tasks}, using tuned and standardized versions. Next, we explore three challenging features of traffic data—anomalies, missing values, and external variation—and evaluate model robustness by measuring performance decline to address \textbf{RQ1}. The evaluation metrics include the Normalized Root Squared Error (NRMSE) 
\footnote{We evaluate our results using other error metrics: mean absolute error, mean squared error, and root mean squared error. All results we discuss using NRMSE hold for these metrics; we show NRMSE for simplicity and omit the others for brevity}, which accounts for data variability and allows comparison across datasets with different scales, the skill score, which measures each model's improvement over a naive baseline, and the difference in NRMSE to evaluate model degradation.

\begin{align}
    MSE &= \frac{1}{N}\sum_{i = 1}^{N}(\hat{y_i} - y_i)^2 \\
    RMSE &= \sqrt{MSE} \\
    NRMSE &= \frac{RMSE}{y_{max} - y_{min}}, \label{eqn:error} \\ 
    \Delta NRMSE &= \frac{NRMSE_{ideal} - NRMSE_{case}}{NRMSE_{real}} \times 100, \label{eqn:error_delta} \\ 
    Skill\;Score &= 1 - \frac{MSE_{forecast}}{MSE_{Naive}}
    \label{eqn:skillscore}
\end{align}

In the second objective, we evaluate the data efficiency of each model across different data sizes by measuring performance using the NRMSE. This addresses \textbf{RQ2} and provides insights into model data efficiency for network traffic prediction, where data collection can be difficult for larger timescales. Additionally, it sheds light on whether the performance of advanced models is due to their deep or data-heavy nature.

Finally, we evaluate the computational efficiency of deep forecasting models, focusing on training time, model size, and energy consumption to address	\textbf{RQ3}. Computational efficiency is essential for model deployment, especially given the limited resources of network devices. Moreover, due to varying network dynamics, biweekly or more frequent model retraining is necessary, making energy-efficient models preferable to minimize operational costs. We measure the training time, GPU power consumption \(P\) per epoch \(i\) and total energy consumption per training \(E\) per equations \ref{eqn:energy_i} and \ref{eqn:energy_total}, and total training cost. We also record the model size on disk. GPU power is recorded using the NVIDIA management libray\cite{pynvml2023} and the tool further developed in \cite{llmpower}.

\begin{align}
    E_i &= P_i \times t_i , \label{eqn:energy_i} \\
    E_{total} &= \sum_{i = 1}^N P_i \times t_i \label{eqn:energy_total}
\end{align}

We use non-parametric statistical tests, which are robust to non-normal error distributions and appropriate for paired comparisons across datasets. Wilcoxon signed-rank tests are used for pairwise comparisons between each model and the reference across datasets, while Nemenyi post-hoc tests are used to evaluate whether model behavior differs significantly across datasets. Holm–Bonferroni correction is applied for multiple testing.
The experimental protocol is summarized in the third section of Fig. 
\ref{fig:methodology}, with subsequent sections detailing each experiment and its findings.

\begin{figure}[t!]
    \centering
    \includegraphics[width=0.95\linewidth]{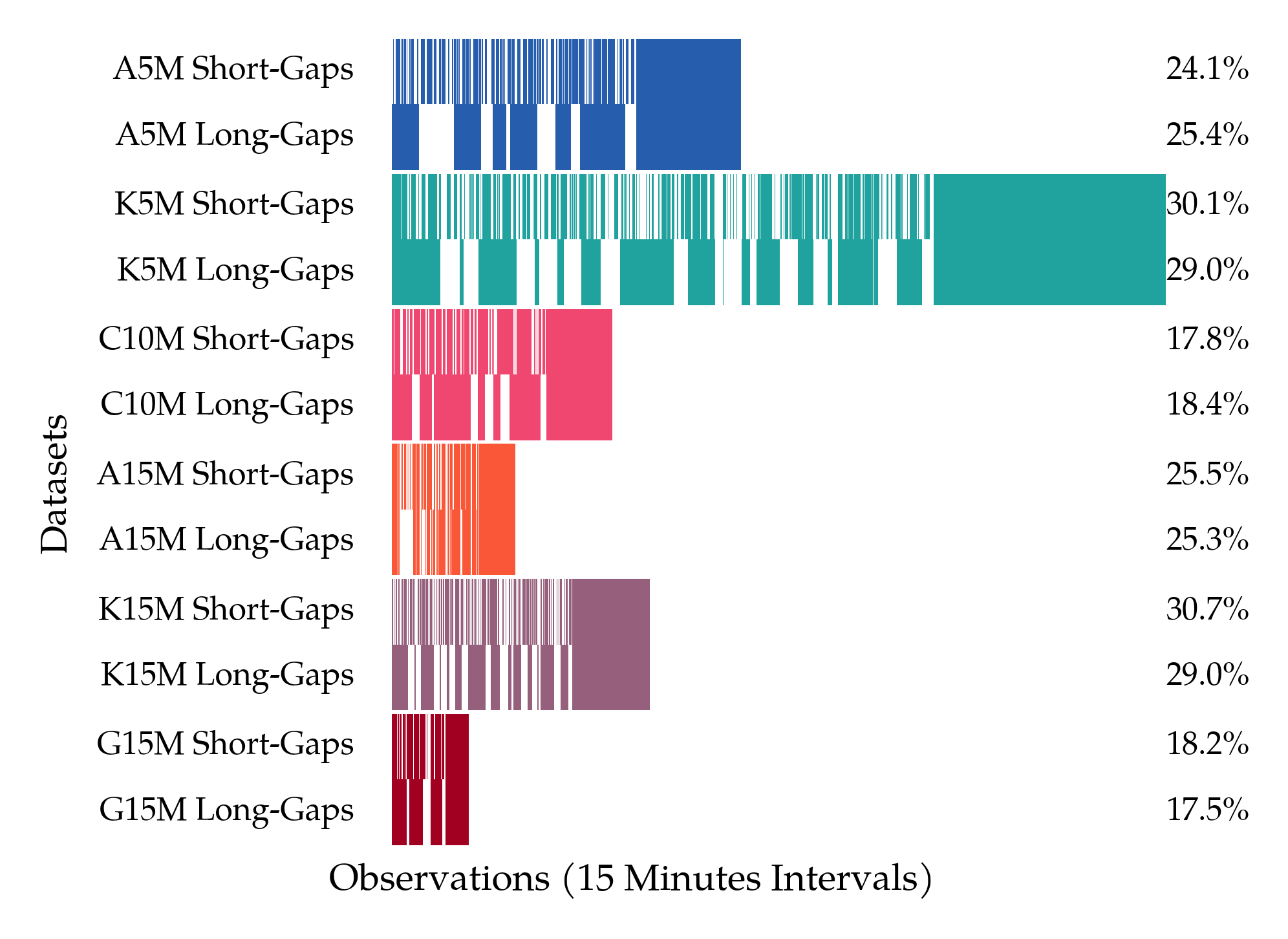}
    \caption{Missingness maps for three dataset pairs illustrate two missing-at-random patterns with different gap durations. Each pair has an upper map showing short-gap missingness (missing intervals up to one day), and a lower map showing long-gap missingness (intervals over several days) week.}
    \label{fig:missingmap}
\end{figure}

\section{Multi-Scale Performance and Robustness across network environments}
\label{sec:exp_generalization&Robustness}


In this section address \textit{RQ1} by evaluating model performance across various time scales and network traffic datasets. We then isolate three challenges-anomalies, missing values, and variations-examining their impact on model robustness and providing insights into the models' resilience to real-world network traffic issues.

\begin{figure*}
    \centering
    \includegraphics[width=\linewidth]{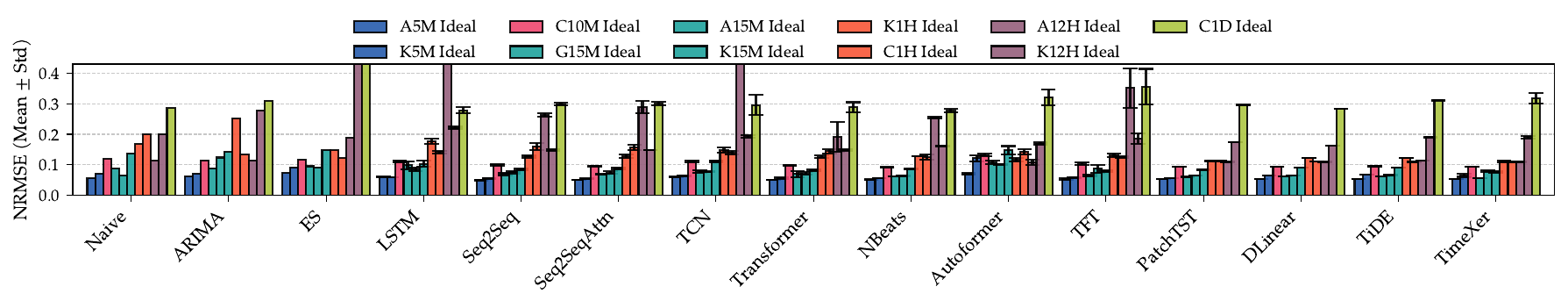}
    \caption{Quantitative evaluation of model accuracy across 11 datasets. Models on the x-axis are arranged in chronological order, and for each model, 11 bars report the NRMSE for each dataset, with error bars indicating the standard deviation across repeated runs. This comparison highlights differences in accuracy among models as well as the degree of performance variability across datasets.}
    \label{fig:exp10_nrmsebars}
\end{figure*}

\begin{figure}[t!]
    \centering
    \includegraphics[width=\linewidth]{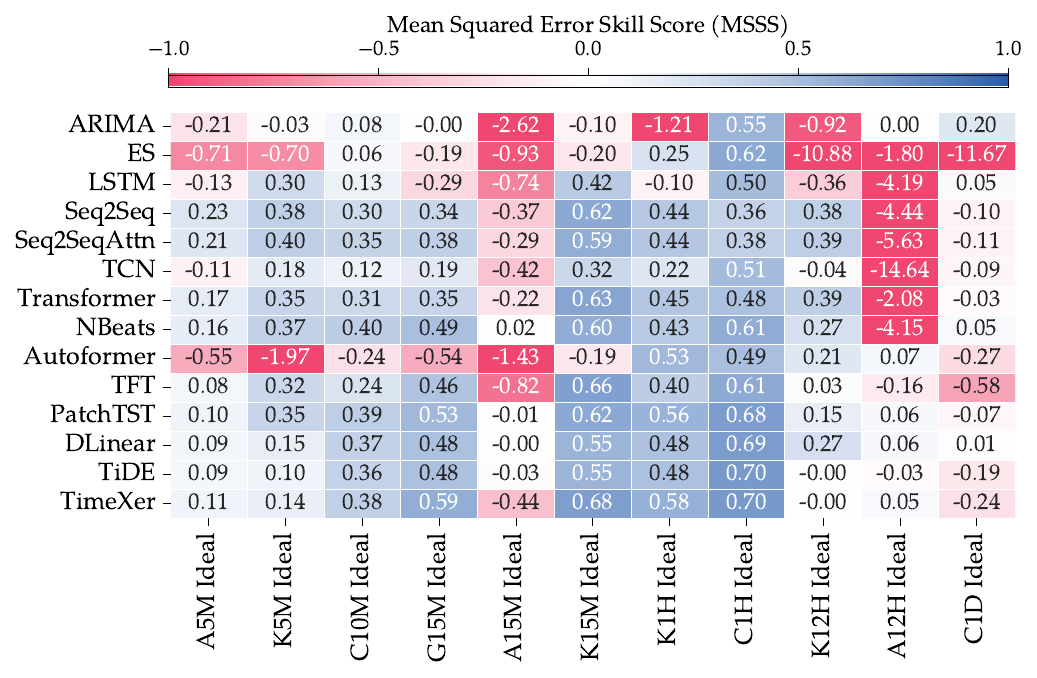}
    \caption{Heat map of model performance across the 11 datasets, where rows represent models and columns represent datasets. Each cell shows the MSE-based skill score, with color intensity reflecting relative performance. The visualization reveals systematic differences between models and highlights variability in skill across datasets.}
    \label{fig:exp10_skillscore}
\end{figure}

\subsection{Evaluation Methodology}
In the first half, we establish the model performance on high-quality ideal data by scaling and cleaning the datasets described in Table \ref{tab:ntp_tasks}. Specifically, we filter outliers with the Three-sigma rule and replace them with forward fillings, as these techniques were determined best \cite{saha2023outliers}. Additionally, we impute large data gaps using seasonal-trend decomposition with interpolation to preserve seasonal patterns. This involves decomposing the traffic series into trend and seasonal components, interpolating each one, and then reconstructing the series. We subscript the filtered and imputed datasets \textit{Ideal}.

For the second half of the experiment, we examine real performance by isolating each traffic characteristic as follows.

\subsubsection{Anomlies} 
To analyse the impact of anomalies, we impute the short- and medium-timescale datasets without filtering anomalies, resulting in \textit{A5M\_Outlier, K5M\_Outlier, C10M\_Outlier, }and \textit{G15M\_Outlier}, and train a model version for each. 



\subsubsection{Missing Values}
Similarly, we analyze the impact of data gaps using the ideal datasets, where anomalous values were replaced by forward-filling and native missing values were imputed with seasonal decomposition and residual interpolation to retain seasonal structure and approximate local variations. This isolates the injected gap patterns from anomalies and pre-existing gaps. We then introduce random gaps represented by zeros in two configurations: short gaps of up to one day and long gaps lasting one week. The resulting missingness ranges from approximately \(17\%\) to \(31\%\), with similar percentages within each short- and long-gap dataset pair, longer datasets exhibiting more gaps to better mimic real-world conditions and assess the effects of different missingness levels and mechanisms. The resulting datasets are shown in Fig.~\ref{fig:missingmap}.

\subsubsection{Extrenal Factors} 

Lastly, we evaluate model sensitivity to external traffic variations through calendar and weather-related factors. For CESNET, the available external context consists of calendar information, and we evaluate the effect of holidays. For the Campus dataset, which was collected and monitored locally, we additionally evaluate weather-related variations using wind speed and rainfall. For the six models that incorporate exogenous variables, holidays are modeled as a binary variable where \(1\) indicates a holiday. Weather variables are processed and scaled from \cite{windspeed}. We conduct the analysis on hourly-scale datasets, K1H and C1H, where the effects of external factors are most pronounced. We also evaluate performance on abnormal days within \(3\) days before and after the event to measure reaction speed and recovery. 

Across all experiments, each model is trained and tested five times. Data are split into 70\% training, 10\% validation, and 20\% testing. We employ a rolling cross-validation scheme in which models are trained on an expanding window and evaluated on the subsequent fold. The test fold from each step is added to the training set before predicting the next fold. Final performance is reported as the average error (equations \ref{eqn:error}-\ref{eqn:error_delta}) and skill score (equation \ref{eqn:skillscore}) across all folds, along with the corresponding standard deviation to reflect variability across runs.

\subsection{Model Behavior}

\begin{table}[h]
\centering
\caption{Mean skill score vs Naive baseline (percentage improvement) with Wilcoxon significance up to six digits}
\label{tab:skill_score_mean}
\renewcommand{\arraystretch}{1.3} 
\begin{tabularx}{\linewidth}{l >{\centering\arraybackslash}X r c}
\toprule
Model & Mean Skill Score (\%) & $p\_{\text{Wilcoxon}}$ & Significance \\
\midrule
DLinear     &         +31.0\% &    0.000038 &          *** \\
PatchTST    &         +28.8\% &    0.005329 &           ** \\
TiDE        &         +26.6\% &    0.044559 &            * \\
TimeXer     &         +18.5\% &    0.049366 &            * \\
TFT         &         +17.9\% &    0.018082 &            * \\
Transformer &         +17.6\% &    0.022987 &            * \\
NBeats      &         +11.8\% &    0.000336 &          *** \\
Seq2Seq     &          +3.1\% &    0.022987 &            * \\
Seq2SeqAttn &          -3.7\% &    0.022987 &            * \\
LSTM        &         -26.6\% &    0.352459 &         n.s. \\
Autoformer  &         -33.6\% &    0.952984 &         n.s. \\
TCN         &         -67.6\% &    0.293499 &         n.s. \\
ARIMA       &         -41.1\% &    0.743715 &         n.s. \\
ES          &        -147.3\% &    0.828785 &         n.s. \\
\bottomrule
\end{tabularx}
\end{table}

\subsubsection{Ideal Performance} 
\label{subsub:ideal_performance}
Fig. \ref{fig:exp10_nrmsebars} displays individual model performance (NRMSE) across the specified NTP tasks, with error bars representing variability across multiple runs. As the error bars show, recurrent models exhibit substantial run-to-run variability reflecting sensitivity to random weight initialization during recurrent training. Meanwhile, most models become unstable at the coarsest scales (12 hours and 1 day), where temporal aggregation substantially reduces the number of available training observations.

To quantify model performance against the naive baseline, Fig. \ref{fig:exp10_skillscore} and Table \ref{tab:skill_score_mean} present these individual and average skill scores per NTP task, respectively. We observe three performance regimes that reflect the interaction between the traffic dynamics and model processing mechanisms:

\noindent\textit{At short timescales (5 minutes)}, traffic retains strong local fluctuations and rapidly changing dependencies. Seq2Seq models perform well in this regime because their recurrent encoder--decoder structure summarizes recent temporal information to guide the decoding of the forecast sequence, making them effective at tracking short-term traffic dynamics despite their lower stability.

\noindent\textit{At medium scales (\(10\)--\(15\) minutes),} aggregation suppresses part of the high-frequency variation while preserving local fluctuations and making periodic structure more pronounced. Patch-based models benefit from this mixture; both PatchTST and TimeXer preserve local patterns within temporal patches while attention captures dependencies over longer ranges. Also, NBeats performs strongly because its stacked basis-expansion blocks can represent both nonlinear short-term behavior and emerging trend--seasonal structure. These properties explain why patching and NBeats become more competitive as traffic transitions from highly irregular fine-scale behavior toward clearer periodic organization.

\noindent\textit{At hourly scales}, local variations are smoothed with prevalent daily and weekly structure (identified in Fig.\ref{fig:frequency}). DLinear's trend--seasonal separation and linear mappings, TiDE's dense representation with a direct linear residual path, and patch-based attention provide better ways to model this structure without recurrent processing of every observation. However, at the coarsest scales, 12 hours and 1 day, the reduced number of observations is accompanied by a month-long forecasting horizon. Several models then have small or negative skill relative to the naive baseline.

We rank the models' ideal performance in Table \ref{tab:skill_score_mean} by their average skill scores across all tasks along with their statistical significance. DLinear and PatchTST lead, outperforming the baseline in the majority of tasks, followed by TiDE and TimeXer. LSTM, Seq2SeqAttn, TCN, and Autoformer show negative mean skill scores because they perform poorly on specific tasks.

\noindent\underline{\textbf{\textit{Insight \#1.}}}
{Deep forecasting model performance depends on the time scale of traffic data. Sequential models achieve peak accuracy at short timescales, but their superiority diminishes as fast dynamics are smoothed out and periodic patterns emerge. Conversely, linear, MLP-based models and patching transformer models outperform the baseline across the majority of tasks, remaining effective across multiple time scales, making them suitable for pipelines requiring diverse NTP tasks.}

\begin{table}
\centering
\footnotesize
\setlength{\tabcolsep}{2pt}
\renewcommand{\arraystretch}{1.05}
\newcommand{\hpocell}[2]{\shortstack{$#1$\\[-1pt]{\scriptsize $#2$}}}
\caption{Practical tunability and observed HPO effort across eight tasks. Tunability: mean with $\pm$SD beneath; HPO quantities: median with [Q1, Q3] beneath.}
\label{tab:tuning_cost}
\begin{tabular*}{\columnwidth}{@{\extracolsep{\fill}}lcccc@{}}
\toprule
\textbf{Model} &
\shortstack{\textbf{Tunability}\\\textbf{(\%)}} &
\shortstack{\textbf{Evaluated}\\\textbf{configs.}} &
\shortstack{\textbf{HPO time}\\\textbf{(min)}} &
\shortstack{\textbf{HPO cost}\\\textbf{(min/config.)}} \\
\midrule
LSTM & \hpocell{9.18}{\pm 11.75} & \hpocell{36.5}{[23.5,50.0]} & \hpocell{60.13}{[39.93,61.34]} & \hpocell{1.73}{[0.80,3.14]} \\
\addlinespace[2pt]
Seq2Seq & \hpocell{-1.76}{\pm 5.22} & \hpocell{18.0}{[9.75,37.25]} & \hpocell{63.55}{[60.72,65.87]} & \hpocell{3.87}{[1.65,9.34]} \\
\addlinespace[2pt]
Seq2SeqAttn & \hpocell{1.72}{\pm 7.77} & \hpocell{28.0}{[10.0,41.75]} & \hpocell{62.29}{[57.28,65.41]} & \hpocell{2.39}{[1.40,8.36]} \\
\addlinespace[2pt]
TCN & \hpocell{0.99}{\pm 22.10} & \hpocell{7.0}{[5.5,15.5]} & \hpocell{67.70}{[61.64,73.75]} & \hpocell{9.42}{[4.14,14.15]} \\
\addlinespace[2pt]
Transformer & \hpocell{-0.59}{\pm 8.30} & \hpocell{5.5}{[3.75,10.25]} & \hpocell{65.23}{[61.14,69.81]} & \hpocell{12.27}{[6.41,18.18]} \\
\addlinespace[2pt]
NBeats & \hpocell{1.26}{\pm 2.68} & \hpocell{10.0}{[3.0,18.5]} & \hpocell{62.18}{[60.72,64.35]} & \hpocell{3.81}{[2.40,15.08]} \\
\addlinespace[2pt]
Autoformer & \hpocell{-12.57}{\pm 34.04} & \hpocell{5.0}{[3.0,5.5]} & \hpocell{70.61}{[66.17,75.53]} & \hpocell{14.80}{[12.75,33.78]} \\
\addlinespace[2pt]
TFT & \hpocell{-4.32}{\pm 10.01} & \hpocell{3.0}{[1.75,4.25]} & \hpocell{79.75}{[69.23,97.32]} & \hpocell{32.52}{[15.85,60.00]} \\
\addlinespace[2pt]
PatchTST & \hpocell{16.98}{\pm 10.35} & \hpocell{36.0}{[22.5,48.5]} & \hpocell{60.85}{[56.76,62.38]} & \hpocell{1.77}{[1.19,2.81]} \\
\addlinespace[2pt]
DLinear & \hpocell{1.01}{\pm 0.94} & \hpocell{50.0}{[39.5,50.0]} & \hpocell{27.33}{[20.50,60.79]} & \hpocell{0.55}{[0.41,1.88]} \\
\addlinespace[2pt]
TiDE & \hpocell{3.58}{\pm 4.07} & \hpocell{40.0}{[24.25,50.0]} & \hpocell{57.77}{[45.98,61.70]} & \hpocell{1.60}{[0.92,2.55]} \\
\addlinespace[2pt]
TimeXer & \hpocell{18.42}{\pm 15.90} & \hpocell{13.5}{[5.75,23.0]} & \hpocell{60.60}{[60.34,62.21]} & \hpocell{5.36}{[2.71,11.20]} \\
\addlinespace[2pt]
\bottomrule
\end{tabular*}
\end{table}

\begin{figure}[t!]
    \centering
    \includegraphics[width=\linewidth]{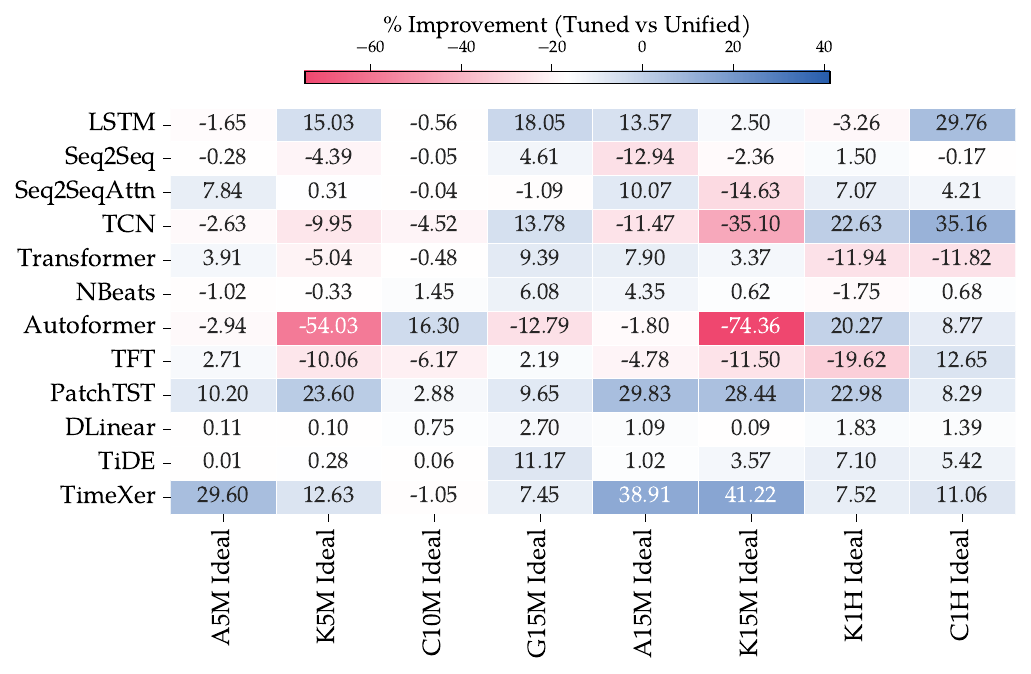}
    \caption{Percentage change in model performance after tuning relative to the unified (untuned) baseline across all datasets. Rows represent models and columns represent datasets, ordered as in the experimental design. Each cell shows (unified-tuned)/unified\(\times\)100\%; cool colors indicate improvement with tuning, warm colors indicate degradation. }
    \label{fig:exp10_standaridez}
\end{figure}

\subsubsection{Tuning Sensitivity and Cost}
Table~\ref{tab:tuning_cost} summarizes each model's practical tunability and HPO effort across eight prediction tasks, while Fig.~\ref{fig:exp10_standaridez} examines their task-specific sensitivity to optimization. 
As in Table~\ref{tab:tuning_cost}, PatchTST and TimeXer obtain the largest average gains, $16.98\%$ and $18.42\%$, respectively. This indicates that their task-specific configuration (patch size) contributes to their strong tuned performance across most tasks, as further shown in Fig. \ref{fig:exp10_standaridez}. In contrast, DLinear exhibits only $1.01\pm0.94\%$ tunability despite completing a median of $50$ trials, indicating limited reliance on task-specific tuning to achieve its performance. NBeats and TiDE also achieve modest tuning gains.

In contrast, other models exhibit substantial differences in tuning gains across tasks. For instance, recurrent models show mixed benefits; LSTM achieves an average gain of $9.18\pm11.75\%$, while Seq2Seq and Seq2SeqAttn achieve $-1.76\pm5.22\%$ and $1.72\pm7.77\%$, respectively. Notably, TCN has an average gain of $0.99\%\pm22.10\%$, while Fig. \ref{fig:exp10_standaridez} shows tunability ranges between $-35\%$ and $+35\%$, indicating that TCN tuning can substantially improve or degrade performance depending on the task. Likewise, Autoformer and TFT exhibit negative average tunability and considerable task-to-task variation. This is because their high HPO costs of $14.80$ and $32.52$ minutes per configuration, respectively, limit exploration under the common budget, with medians of only $5$ and $3$ configurations evaluated per task.

The common tuning budget constrains computational expenditure while exposing differences in tuning cost and gains across model--task combinations. For network operators, model selection should consider both the observed benefit of tuning and the effort required to obtain it, particularly when models must be retuned across traffic settings.

\noindent\underline{\textbf{\textit{Insight \#2.}}}
Tuning requirements differ substantially across model families. Patch-based models show large gains from task-specific optimization, whereas linear and MLP models show more modest gains under the common tuning budget. Therefore, model selection should consider the accuracy achieved across the required time scales together with the tuning effort needed across traffic settings.

\begin{figure}
    \centering
    \includegraphics[width=\linewidth]{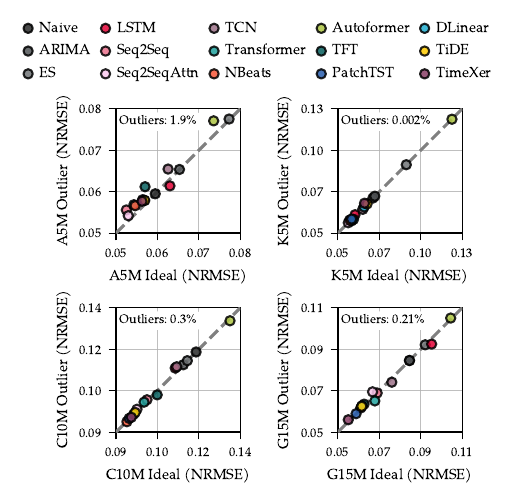}
    \caption{Scatter-plot comparison of model error across four datasets. Each subplot displays the error computed using datasets with outliers versus that computed using outlier-filtered datasets. The identity line represents perfect equivalence between the two conditions, with points closer to the line indicating reduced sensitivity of the error metric to outliers.}
    \label{fig:exp11_anomalies}
\end{figure}

\subsubsection{Robustness to Outliers}
Fig. \ref{fig:exp11_anomalies} shows relative model performance on datasets with anomalies. Statistical models demonstrate robustness to outliers across all datasets. In contrast, deep learning models' performance declines in proportion to outlier percentage; datasets with more than \(0.3\%\) outliers exhibit an average error increase of \(3.5\%\) in deep forecasting models. Low-capacity models such as TiDE and DLinear exhibit only \(1.75\%\) degradation, whereas complex models like TFT experience a \(7.5\%\) performance decline. This contrast is consistent with the scale-dependent model strengths observed under ideal conditions. Recurrent architectures that benefit from fine-scale local dynamics are also more sensitive when those observations are corrupted, whereas simpler models that perform well on smoother traffic regimes exhibit greater robustness to outliers. Specifically, DLinear's decomposition and linear mappings emphasize broader temporal structure, and TiDE's direct linear residual path retains this structure alongside nonlinear modeling, which helps explain their lower observed degradation. Surprisingly, LSTM slightly improves on A5M and G15M, suggesting that some retained outliers may contain useful traffic information that the model can exploit.

Table \ref{tab:anomlies_significance} further reports the statistical significance of collective performance across these datasets. It distinguishes A5M, which contains the highest proportion of outliers (1.9\%), from the three lower-outlier datasets. Its robustness behavior differs significantly from K5M, C10M, and G15M ($p\leq0.01$), while the latter three datasets show no significant differences. 
These results extend prior findings by Saha et al. \cite{saha2023outliers}, which evaluated sequence models on a single dataset with a \(1\%\) outlier percentage. 

\noindent\underline{\textbf{\textit{Insight \#3.}}}
Model performance degrades as the proportion of outliers increases. Simple, low-capacity models exhibit greater robustness than RNNs under high outlier rates, highlighting the importance of explicit outlier handling when deploying sequential models, or alternatively choosing simpler models at comparable error levels.

\begin{table}
\centering
\caption{Statistical Significance of model robustness to outliers per data set using the Friedman-Nemenyi test}
\label{tab:anomlies_significance}
\renewcommand{\arraystretch}{1.3} 
\begin{tabularx}{\linewidth}{l 
>{\centering\arraybackslash}X 
>{\centering\arraybackslash}X 
>{\centering\arraybackslash}X 
>{\centering\arraybackslash}X 
}
\toprule
 & a5m & k5m& c10m & g15m \\
\midrule
a5m & 1.0 n.s. & 0.006 ** & 0.006 ** & 0.01 * \\
k5m & 0.006 ** & 1.0 n.s. & 1.0 n.s. & 0.999 n.s. \\
c10m & 0.006 ** & 1.0 n.s. & 1.0 n.s. & 0.999 n.s. \\
g15m & 0.01 * & 0.999 n.s. & 0.999 n.s. & 1.0 n.s. \\

\bottomrule
\end{tabularx}
\end{table}

\begin{figure*}
    \centering
    \includegraphics[width=0.99\linewidth]{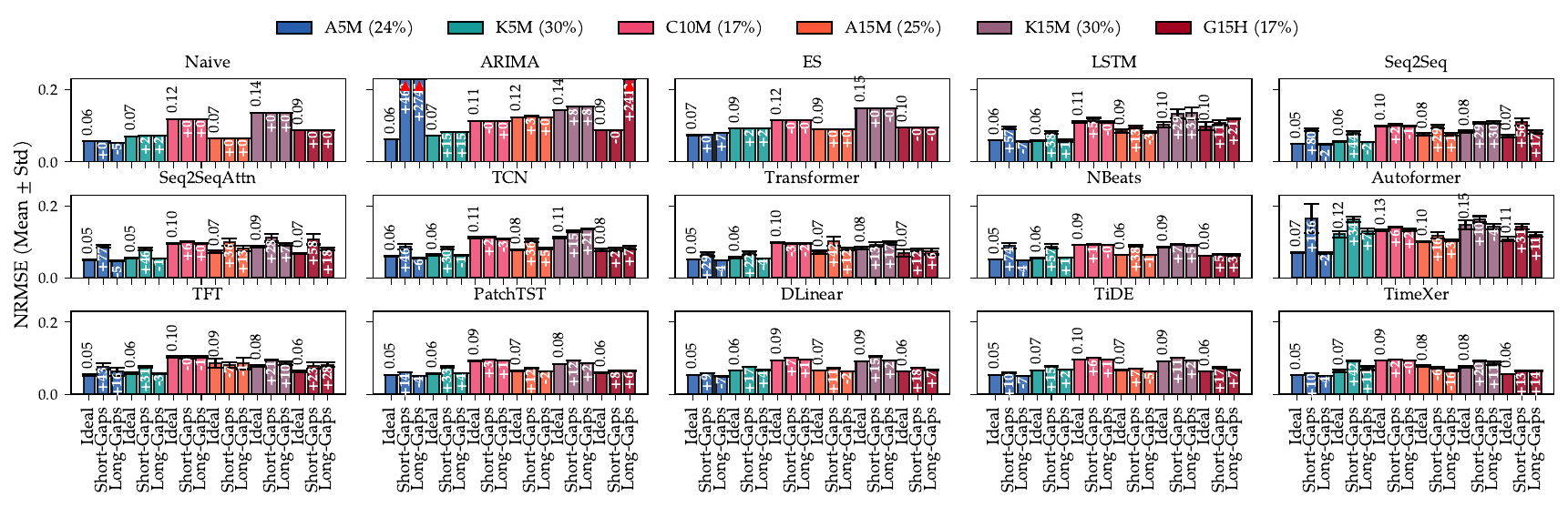}
    \caption{Missingness impact on model performance across 18 datasets. Each subplot corresponds to a model, with bars grouped by dataset type: ideal (no missing values), short-gap (maximum one-day missing intervals), and long-gap (missing intervals up to one week). NRMSE values are shown above the original bars, while vertical white annotations inside short-gap and long-gap bars indicate the percentage change relative to the original. Overshooting bars are capped at the plotted y-axis limit, with arrows indicating truncation.}
    \label{fig:exp12_missing}
\end{figure*}

\subsubsection{Robustness to Missing Data}
Fig. \ref{fig:exp12_missing} compares the robustness of models to missing data across different time scales and missingness patterns, specifically long and short gaps. Overall, the nature of data gaps has a greater influence than the percentage of missing data across all datasets and time scales. 
On short timescales (5 minutes), short gaps have approximately \(20\times\) the impact of long gaps on model performance; on medium timescales, the impact of short gaps is about \(10\times\) on average. 
At a fixed missingness percentage, distributing gaps across more locations can interrupt more input--output windows and repeated portions of the daily cycle, while concentrating missingness in a longer interval leaves longer contiguous observations elsewhere.

Sequential models and NBeats are most affected by missing data, struggling to learn short-term dependencies. Conversely, Dlinear and TiDE models demonstrate higher robustness, with an average degradation of \(17\%\), compared to \(76\%\) for Sequential models. 
This is because the zero-filled gaps in this experiment introduce abrupt transitions into the observed sequences, which affect the local temporal information that benefits sequential models on the ideal fine-scale data. The lower degradation of DLinear and TiDE is consistent with their ability to retain broader structure through decomposition or a direct linear path. Thus, the fine-scale accuracy advantage identified earlier must be considered together with the continuity of the telemetry available for training and prediction.

\noindent\underline{\textbf{\textit{Insight \#4.}}} The characteristics of data gaps and time scale are more critical than missingness percentage alone in model development. Under the same percentage and randomness mechanism, short gaps of up to one day have a greater impact on model performance than longer gaps, particularly at short scales. This indicates that deep forecasting models are more sensitive to local information loss, which is advantageous since short gaps are relatively easier to impute.

\begin{figure*}[t!]
    \centering
    \begin{minipage}[t]{0.7\textwidth}
        \centering
        \includegraphics[width=\linewidth]{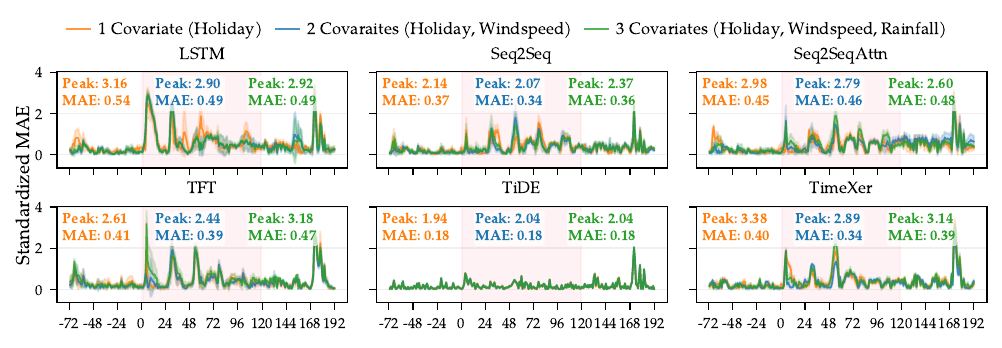}\par
        \vspace{0.4ex}
        \includegraphics[width=\linewidth]{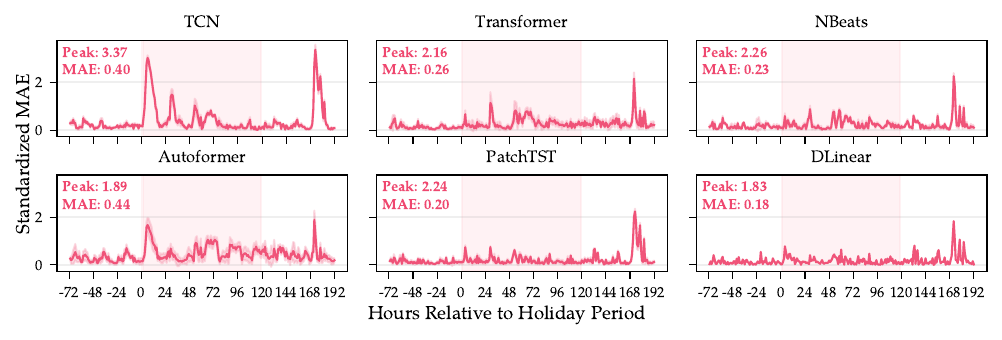}\par
        \vspace{0.6ex}
        \centerline{\footnotesize (a)}
        \label{subfig:exp1.3_behaviour}
    \end{minipage}
    \hfill
    \begin{minipage}[t]{0.29\textwidth}
        \centering
        \includegraphics[width=\linewidth]{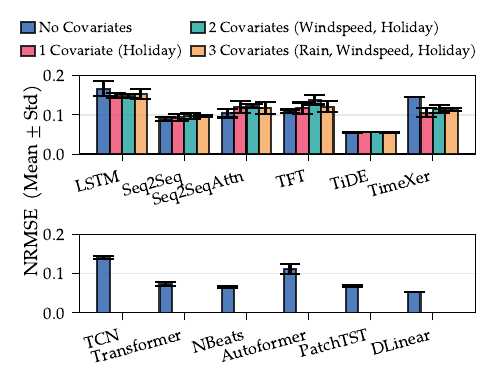}\par
        \vspace{2.3ex}
        \includegraphics[width=\linewidth]{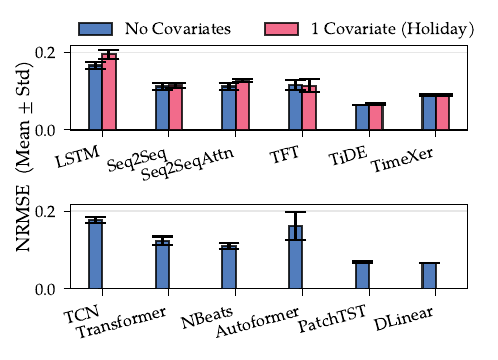}\par
        \vspace{1.6ex}
        \vfill
        \centerline{\footnotesize (b) }
        \label{subfig:exp1.3_bars}
    \end{minipage}
    \caption{
    Model behavior and performance comparison for abnormal days on datasets K1H and C1H across covariates. (a) Model behavior during holidays on K1H, showing absolute error in predicting holiday periods with three covariate configurations: (top row) holiday, holiday + wind, holiday + wind + rain. Each line plot shows the absolute error, with shaded regions representing the standard deviation. The holiday period is shaded for reference. The bottom row shows performance for models without covariates across the same configurations. (b)  Model performance in predicting abnormal days on K1H (top pair) and C1H (bottom pair) with covariates (holiday, wind, rain) and without covariates on dataset C. Error bars show ±1 standard deviation.}
    \label{fig:exp13_combined}
\end{figure*}

\subsubsection{External Factors and Recovery}
Finally, Fig. \ref{fig:exp13_combined} displays model behavior (a) and overall performance (b) for abnormal days (holiday or extreme weather) on an hourly timescale, with and without covariates, based on the model support. In (a), the model behavior for k1H is shown, considering three days before and after abnormal days. Generally, models with or without covariates struggle to recover by the third day—Monday—and continue to predict low traffic. 

Predictors must infer the return to usual traffic patterns from recent history dominated by lower demand, as shown by the abnormal weakening of daily and weekly activity in Fig.~\ref{fig:frequency}, which makes recovery difficult. Moreover, a binary holiday flag identifies the event but does not describe its recovery, whereas wind speed provides a continuous signal negatively correlated with traffic (Fig.~\ref{fig:background_trafficCharacteristics}(e)). This also explains why LSTM, Seq2Seq, TFT, and TimeXer benefit from combining holiday and wind-speed inputs, while adding rainfall peaks offers little additional benefit or even weakens performance. Overall, TimeXer's average error decreases by $15\%$, consistent with its separate representations of target patches and exogenous series. Nevertheless, benefits depend on the model and error criterion; TiDE changes little across configurations, while Seq2SeqAttn's peak error decreases at the expense of slightly higher average error.

Strong performance does not necessarily require external covariates. DLinear achieves the lowest average error and smallest peak in the illustrated window because it predicts an average pattern; PatchTST, N-BEATS and Transformer maintain relatively low average errors but exhibit isolated spikes, whereas TCN and Autoformer show larger, more persistent errors. Hence, for holiday-related capacity or energy scheduling, these results highlight the need to assess the incremental value of contextual inputs alongside event-window accuracy and recovery-period error peaks.

\noindent\underline{\textbf{\textit{Insight \#5.}}} Models with and without exogenous variables respond similarly to abnormal days, particularly in recovering from low traffic periods. While current models effectively combine heterogeneous time series, they do not explicitly model lagged or causal relationships. Future research should focus on integrating causal relations for exogenous variables within NTP models.   

\subsection{Network Management Implications}
\begin{enumerate}[label=\arabic*.]
\item Model choice should follow the task's temporal resolution. For fine-scale load prediction used in routing or congestion management, sequential models are strong candidates. For tasks dominated by daily and weekly demand patterns, operators should consider lightweight MLPs, decomposition models, and patch-based models. 

\item Given the need for continuous tuning, model selection should weigh both tuning gains and the effort required to achieve them. Patch-based models benefit strongly from task-specific optimization, whereas Linear and MLP models yield modest gains with lower tuning effort.

\item When a forecast serves as the expected-traffic reference for anomaly detection, robustness to outliers in the training history is a separate selection criterion from ideal-data accuracy. DLinear and TiDE show more robustness to outliers than sequential models.

\item For intermittently collected telemetry, operators should consider the duration and distribution of gaps as well as the missingness percentage. Preserving the continuity of traffic observations should be prioritized over fixing prolonged gaps due to equipment failures.

\item For capacity planning and energy scheduling around holidays and abnormal days, model selection should consider whether the available external information is sufficient to improve prediction during both the event and its recovery, so as not to risk provisioning errors. Covariate support alone is an incomplete selection criterion.
\end{enumerate}

\section{Traffic Data Efficiency}
\label{sec:exp_dataeffeciency}

We address \textit{RQ2} by comparing model performance in terms of \textit{information quantity} across multiple scales. 
Information quantity\cite{benidis2022deep} measures structured time series like traffic data in periods rather than data points. Collecting sufficient traffic data for coarser scales is more costly than for finer ones, while training on smaller datasets is more resource-efficient. Furthermore, data efficiency evaluates whether the accuracy advantages identified in the previous section persist when fewer repetitions of the daily and weekly traffic patterns are available.

\subsection{Evaluation Methodology}
We utilize five Ideal datasets—A5M, K5M, G15M, k1H, and C1H—imputed and cleaned to ensure data quality and to examine how information quantity varies across scales. To mimic real data collection, we consistently measure durations in weeks across timescales. One week of hourly data contains \(168\) data points but retains comparable period and trend information as it covers the same calendar interval as one week of 5-minute data. Differences in the number of points may arise from input lengths.
Models are trained starting with \(12\) weeks of data -the size typically used in NTP studies- halving the training sample for larger scales (to \(3\) weeks) and reducing it to \(1\) day for shorter scales. A fixed validation and test period of \(3\) weeks is maintained across all experiments. Each configuration is repeated five times. Model performance is evaluated across datasets and data sizes, with results normalized to the \(12\)-week baseline to assess deviations.

\begin{figure}[t!]
    \centering
    \includegraphics[width=\linewidth]{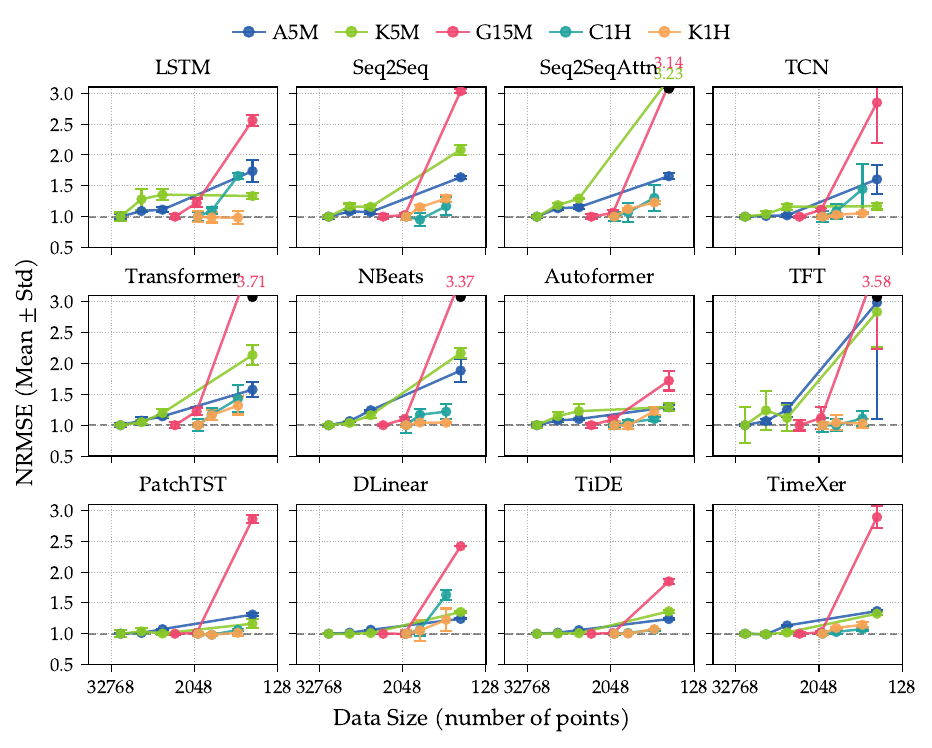}
    \caption{Normalized error (mean ± standard deviation) for all models across dataset sizes. For each model–dataset pair, the error at the largest dataset size is used as the full-data baseline and normalized to 1. Values above 1 indicate degradation as data decreases. The dashed line marks the baseline.}
    \label{fig:exp2_datasize}
\end{figure}

\subsection{Model Behavior}
Fig \ref{fig:exp2_datasize} shows the individual performance of each model across the five datasets with varying data sizes, each normalized to the performance on the largest dataset. 
Models vary in data efficiency, and their reliance on information quantity is largly consistent across timescales and network environments. We observe three sets of behaviors. First, Recurrent models, TFT, NBeats, and pointwise attention models like vanilla Transformer and Seq2Seq\_Attn, are more data-reliant, with up to a \(50\%\) error increase at \(3\) weeks of data and more than \(2\times\) below that.

Second, convolutional TCN shows more stable behavior with a \(20\%\) decline at \(3\) weeks of data, and greater instability at smaller sizes. Lastly, patch-based models, DLinear and TiDE, have the highest data efficiency with an average error increase of only \(6\%\) at \(3\) weeks of training data. 

With the shortest 15-minute training history, one day cannot capture the multi-seasonality components, and overall performance deteriorates. Notably, the differences between model groups also qualify the accuracy results in the previous section. DLinear's simple component mappings and TiDE's direct linear path provide a more constrained route to recurring patterns, while Autoformer's decomposition and sub-series-level Auto-Correlation explicitly emphasize recurring temporal structure. Patch-based models similarly reduce dependence on individual time points by operating on local segments rather than separate tokens for every observation. These mechanisms explain the smaller losses observed for these models. However, the data-efficiency advantage does not extend to all MLP architectures since NBeats performs strongly with ample data but is more sensitive to reduced history. In contrast, the benefit of detailed sequential or point-wise representations is more dependent on having enough observations to estimate their temporal relationships.

\noindent\underline{\textbf{\textit{Insight \#6.}}}
The performance of sequential, deep, and point-wise attention models depends on data volume, making them less suitable for deployment scenarios with limited data. In contrast, DLinear, TiDE, and advanced transformers are up to \(3\times\) more efficient. 

\noindent\underline{\textbf{\textit{Insight \#7.}}}
Three weeks represent the minimum \textit{information quantity} required by data-efficient deep learning models to reliably predict network traffic, serving as a benchmark for evaluating future data-efficient prediction models.

\subsection{Network Management Implications}
\begin{enumerate}[label=\arabic*., start =6]
    \item When a network has only a short usable traffic history, DLinear, TiDE, and patch-based models are strong candidates because they retain more of their full-data accuracy, as opposed to Sequential models and NBeats.
    \item A three-week training history is a useful tested operating point, as the data-efficient models incur an average error increase of about $6\%$ relative to twelve weeks while using one quarter of the collection span. This supports evaluating shorter windows for frequent retraining.
\end{enumerate}

\section{Deployment Resource Efficiency}
\label{sec:exp_resourceEfficiency}

\begin{figure*}
    \centering
    \includegraphics[width=0.98\linewidth]{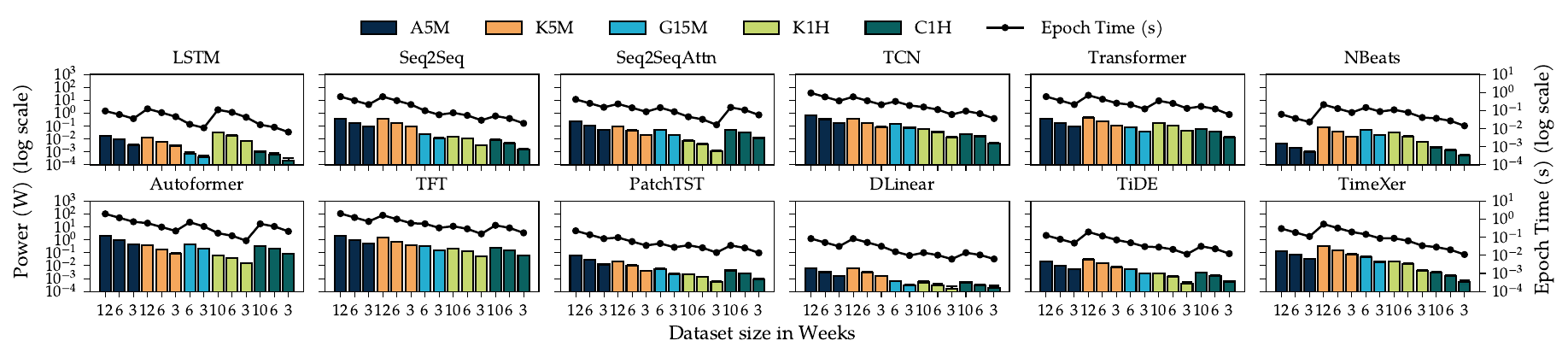}
    \caption{Model Efficiency comparison across multiple data sizes.
    Each of the 12 subplots shows the relationship between power (left y-axis) and time (right y-axis) across 5 distinct data groups (A5M, K5M, G15M, k1H, and C1H). Each group consists of three data sizes (x-axis). The variation in power consumption and per-epoch execution time is analyzed for each group and size combination, highlighting key performance trends and the impact of different time scales. }
    \label{fig:exp3_ResourcperEpoch}
\end{figure*}


The preceding sections show how forecasting accuracy depends on traffic structure, data quality, and training history. We now address \textit{RQ3} by evaluating the computational efficiency of deep forecasting models in terms of training time, energy consumption, and stored model size. Limited network resources and frequent retraining needs—often weekly or biweekly—make these costs important for deployment.

\subsection{Evaluation Methodology}
We measure training resources across the predefined dataset sizes for six datasets—\textit{A5M, K5M, C10M, G15M, K1H, and C1H}—covering short, medium, and long timescales. For each model-data pair, we record the per-epoch training time and average power consumption by sampling the GPU power draw during the epoch. Then, the total training energy consumption is calculated according to equation 
\ref{eqn:energy_total}. Additionally, we record the total model size stored on disk, which serves as a model-agnostic proxy for the memory footprint and reflects model complexity across the diverse architectures evaluated. Each model-dataset pair is repeated five times, and we report the average results. 
We then compare the full-data skill score (equation \ref{eqn:skillscore}) against training time, training energy, and stored model size in separate Pareto fronts. In each plot, a model is considered Pareto-optimal if no other model has at least as high a skill score and at most as much of the selected resource, with a strict improvement in at least one of the two.

\subsection{Model Behavior}

\subsubsection{Training Resource} Fig. \ref{fig:exp3_ResourcperEpoch} depicts each model's power consumption and per-epoch training time across data sizes and time scales. Reducing the training history reduces the number of windows processed per epoch and generally shortens training time and power consumption. As a three-week history processes data from one quarter of the twelve-week collection span, it consumes considerably lower resources. Overall, fine-scale datasets incur higher training costs because the same calendar span contains more observations and training windows. Differences in the selected model configurations also affect the measured costs, as illustrated by NBeats, which uses a smaller model at fine scales to capture traffic fluctuations.

In general, DLinear exhibits particularly low resource consumption and is \(10\times\) more efficient than the LSTM model on average.
Additionally, patching models are the most efficient transformer variant, offering \(10\times\) the efficiency of alternatives. The lower costs of DLinear and patch-based models are consistent with their processing mechanisms.
Surprisingly, TCN is \(10\times\) slower than LSTM under the evaluated configurations, despite its parallelizable architecture, because its selected configuration has higher computational complexity. This also suggests that LSTM effectively captures the challenging dynamics of short-scale traffic data, while other models require significantly more resources to achieve similar performance.
Hence, the TCN--LSTM comparison shows why architectural parallelism alone does not determine measured runtime. Collectively with the previous section, these results make reduced-history training a practical option for models that retain accuracy with fewer observed cycles.

\begin{figure*}
    \centering
    \includegraphics[width=0.98\linewidth]{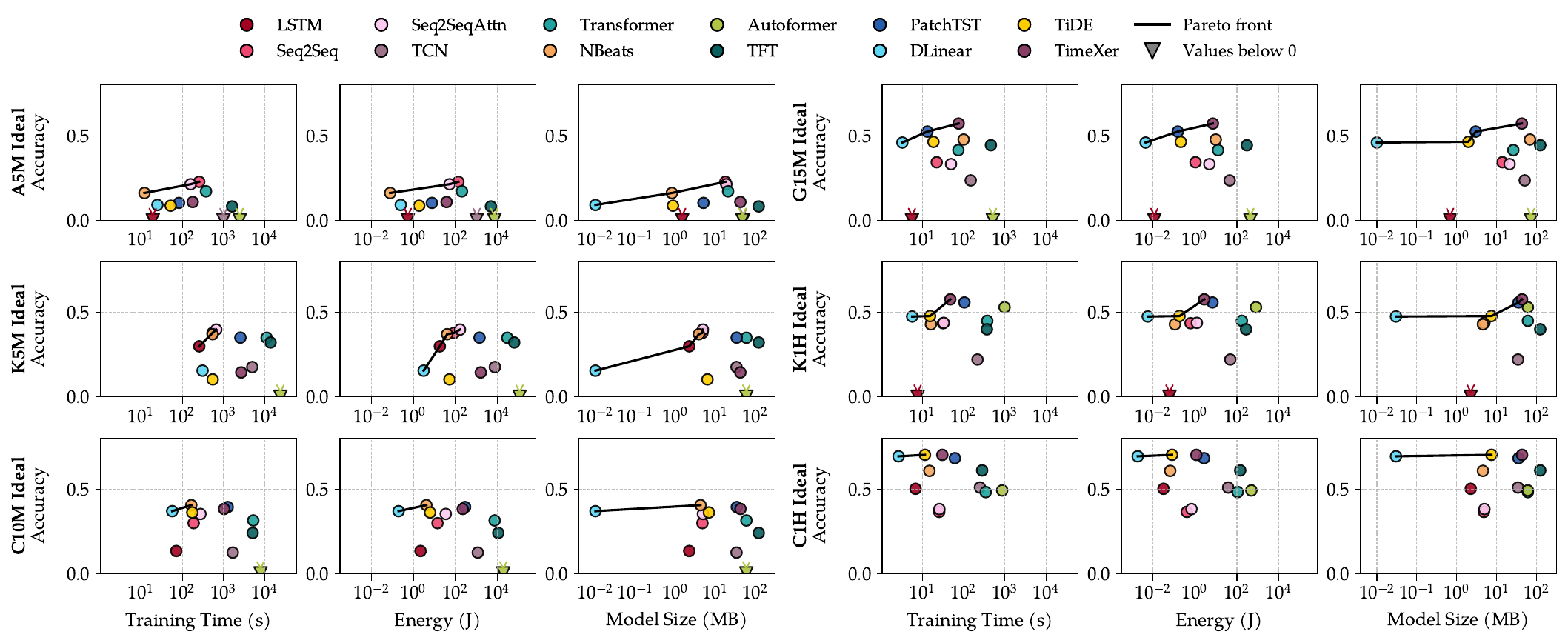}
    \caption{Performance–efficiency trade-offs across datasets and efficiency metrics. Each row corresponds to a dataset, and each column corresponds to an efficiency metric (training time, model size, and energy consumption). Within each subplot, model performance is shown on the y-axis and the corresponding efficiency metric on the x-axis. The Pareto-optimal models for that dataset and metric are highlighted. Performance is evaluated on the full dataset.}
    \label{fig:exp3_paretofront}
\end{figure*}

\subsubsection{Accuracy--Resource Pareto Optimality} Fig. \ref{fig:exp3_paretofront} displays the Pareto front for each accuracy-resource pair across six datasets. Pareto-optimal models are consistent per training resource but vary across traffic timescales.
The change in the models on these fronts follows the accuracy regimes identified in Section~\ref{subsub:ideal_performance}. At fine scales, the pareto front is populated by recurrent models, DLinear, and compact NBeats configurations with small size ( \(<10MB\) ) and energy ( \(<1J\) ) footprint, whereas the trend shifts toward patch-based transformers and TiDE, which become more attractive at medium and hourly scales. Their additional cost --- ( \(<40MB\) ) memory and ( \( <10J\) ) energy consumption--- is useful when it yields an accuracy improvement for the target task. DLinear remains Pareto-optimal in $83\%$ of the evaluated dataset--resource combinations, providing a consistently competitive low-cost option. These full-data fronts complement the data-efficiency requirements; a model on the front must still retain sufficient accuracy at the available history length, while the tuning costs (Table~\ref{tab:tuning_cost}) remain a separate consideration when selecting or retuning its configuration.

\noindent\underline{\textbf{\textit{Insight \#8.}}}
Pareto-Optimal models for network traffic prediction depend strongly on the traffic series time scale. Shifting from Recurrent models, DLinear, and compact NBeats configurations at short scales towards patching-based transformers and TiDE at longer scales.

\noindent \underline{\textbf{\textit{Insight \#9.}}}
Recurrent models, DLinear, and the compact NBeats configuration dominate under tight memory and energy budgets for short time scales, making them better suited to resource-constrained environments, while patching-based transformer models and TiDE incur higher costs.

\noindent \underline{\textbf{\textit{Insight \#10.}}}
Only one model, DLinear, remains Pareto optimal in \(83\%\) of resource and timescale combinations, \(20\times\) smaller and \(10\times\) energy efficient than other models, which makes it highly promising across diverse NTP tasks.


\subsection{Network Management Implications}
\begin{enumerate}[label=\arabic*., start =8]
    \item For fine-scale NTP tasks such as routing or congestion management, recurrent models, DLinear, and compact NBeats configurations provide competitive accuracy--cost trade-offs. The recurrent option is most useful when the local temporal information is available and the training history is sufficient. For hourly capacity planning or energy scheduling, patch-based models and TiDE are useful candidates when their accuracy gains justify the measured additional cost.
    \item Under tight resource budgets, DLinear is a strong initial candidate because it remains Pareto-optimal in $83\%$ of the evaluated combinations and also shows low tuning cost, data efficiency, and robustness to data-quality problems. For tasks needing greater accuracy, sequential, patch-based, or dense encoder--decoder models are worth considering. The deployment choice combines the required timescale and horizon with usable history, telemetry quality, and measured training and model-storage costs.
\end{enumerate}

\section{Which Deep Learner?}
\label{sec:which_deep_learner}
\begin{figure*}
    \centering
    \includegraphics[width=\linewidth]{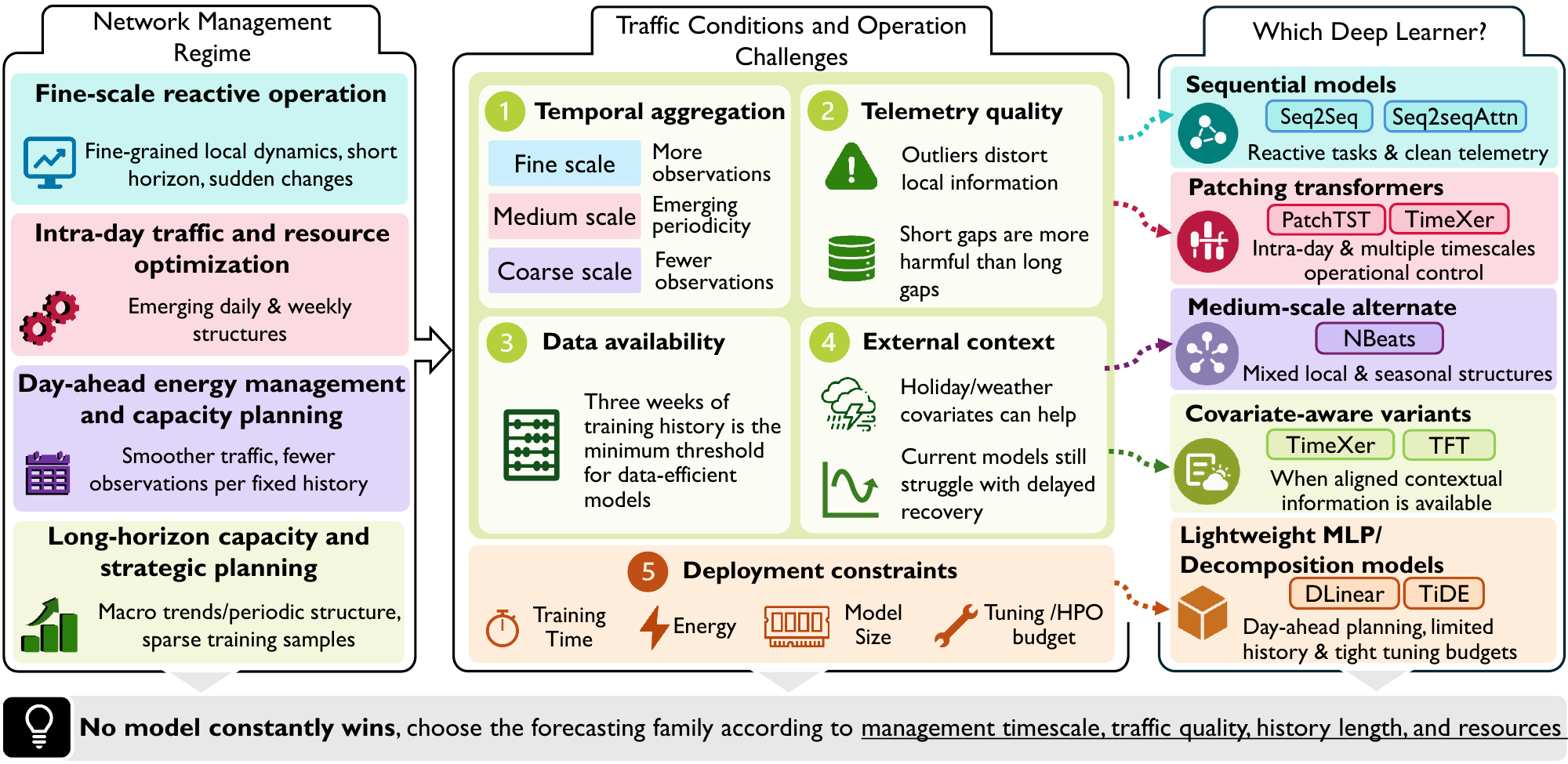}
    \caption{Practical guideline for deep forecasting model selection in network traffic prediction. Network-management tasks are linked to their traffic characteristics, operational challenges, and deployment constraints, leading to conditional model recommendations based on our findings.}
    \label{fig:guidelines}
\end{figure*}

We culminate the findings from the previous sections to guide model selection, as shown in Fig. \ref{fig:guidelines}. For 5-min traffic and short-horizon decisions, recurrent models, particularly Seq2Seq and Seq2SeqAttn, are suitable when accuracy is the priority, provided that anomalies and gaps are treated and sufficient training history is available. NBeats offers a compact alternative when enough history can be retained. As traffic moves towards 10--15-min and hourly scales, PatchTST and TimeXer become attractive choices for exploiting periodic traffic structure, with task-specific tuning included in the deployment budget. TiDE provides a lightweight alternative for hourly, day-ahead forecasting. For month-long forecasts at the coarsest scales, operators should first verify that the selected deep learner improves over the naive baseline, as greater model complexity does not consistently provide an advantage due to limited data at these scales.

When resources, data availability, or telemetry quality constrain deployment, DLinear is a strong starting point because it combines low computational cost, limited reliance on tuning, and robustness to the tested outliers and gaps. TiDE offers another suitable option, while patch-based models remain attractive when limited history is the main constraint and tuning resources are available. For frequent retraining, the observed three-week \textit{information quantity} threshold provides a practical starting point for these data-efficient models. 

For data quality, excessive outliers should be eliminated, and gap handling should account for how missing observations are distributed, since dispersed interruptions can be more damaging than a concentrated outage. For holiday- and weather-related capacity or energy scheduling, TimeXer is worth considering when informative weather inputs are available, but each covariate should be assessed against both average error and recovery-period peaks. Moreover, such schedules require reliable event recovery as well as good overall accuracy.

\section{Limitations }
\label{sec:limitatiation}
Our study focuses on advances in deep forecasting models, which continue to dominate the NTP field. Recently, there has been a trend in the Time Series Forecasting community toward utilizing Large Language Models (LLMs) to develop universal models. Future work will evaluate LLMs with an emphasis on multi-scale generalization. A major drawback of LLMs is their massive resource requirement, which is often undesirable for NTP tasks. It is important to investigate whether these models can achieve generalization gains and whether techniques to reduce their cost can provide acceptable performance-efficiency trade-offs.
Additionally, our evaluation excludes sub-minute time scales, as traffic data at that resolution exhibit different statistical properties and demand specialized analytical treatment. As a result, our findings should not be directly extrapolated to those dynamics. Nonetheless, prior to adopting large transformers or LLM-based models, we establish performance-efficiency baselines and demonstrate what classical, simpler, and advanced architectures can achieve.

\section{Path Forward}
\label{sec:discussion}

Progress in NTP requires forecasting architectures whose representations match traffic dynamics and whose accuracy gains remain attainable under practical data and resource constraints.
Our findings indicate that future NTP deep learning research should prioritize (1) strengthening short-scale MLP models, (2) developing patching and segmentation techniques to better capture short-term traffic dynamics in transformer models, and (3) modeling causal relationships between traffic and external covariates to enable post-event recovery, through improved models or data representations. Solving these challenges while limiting training and tuning costs will balance scalability with competitive accuracy.

\section{Conclusions}
\label{sec:conlusion}

This study empirically evaluates 12 deep forecasting models (8 advanced TSF models and 4 traditional DL architectures) with diverse architectures, processing logic, and innovations against 3 statistical baselines and across more than 20 configurations of 4 real network traffic traces to assess performance across different time scales and prediction horizons, covering a range of traffic prediction tasks. We isolate three key challenges—anomalies, data gaps, and external variations—to evaluate the models' robustness to each. Additionally, we examine data and resource efficiency. Our results offer insights for model deployment and future research directions, highlighting the potential of overlooked MLP architectures and guiding informed exploration of representation techniques for newly emerged transformer models.

\bibliographystyle{IEEEtran}
\bibliography{references}

\end{document}